\documentstyle[epsfig,aps]{revtex}
\begin{document}
\draft
\def\lsim{\lower.5ex\hbox{$\; \buildrel < \over \sim \;$}}
\def\gsim{\lower.5ex\hbox{$\; \buildrel > \over \sim \;$}}
%\setcounter{page}
%\begin{center}
%\begin{bf}
\title{ Scalar and Spinor Perturbation to the Kerr-NUT Space-time \\
%OR\vskip0.5cm
%Study of generalized Kerr-NUT space-time perturbed by scalar and spinor waves \\
%OR\vskip0.5cm
%Klein-Gordan and Dirac equation in generalized Kerr-NUT geometry \\
%OR\vskip0.5cm
%Any Other..............}
}
%\vskip0.5cm
%\vskip0.35cm
%\end{large}
%\end{bf}
%\end{center}
%\noindent\\
\author  {Banibrata Mukhopadhyay\thanks{At present: Harvard-Smithsonian Center for 
Astrophysics, 60 Garden Street, Cambridge, MA 02138, USA; bmukhopa@cfa.harvard.edu} 
\& Naresh Dadhich}
\address{Inter-University Centre for Astronomy and Astrophysics,
Post Bag 4, Ganeshkhind, Pune-411007, India}
%\end{bf}
%\vskip0.5cm
%\noindent\\
%\thanks{e-mail:  bm@prl.ernet.in } \\
%Present address: Physical Research Laboratory, Navrangpura, Ahmedabad-380009,
%India; e-mail: bm@prl.ernet.in}
%\end{center}
\maketitle
\baselineskip = 18 true pt
\vskip0.3cm
\setcounter{page}{1}
%\noindent{Submitted to appear }
\def\ch{\lower-0.55ex\hbox{--}\kern-0.55em{\lower0.15ex\hbox{$h$}}}
\def\lh{\lower-0.55ex\hbox{--}\kern-0.55em{\lower0.15ex\hbox{$\lambda$}}}
%\begin{centre}
%{\bf Abstract}
%\end{center}
%\vskip0.3cm
%\noindent
%{\it Received ...., accepted ........}\\
%\vskip0.3cm

\begin{abstract}

We study the scalar and spinor perturbation, namely the Klein-Gordan and Dirac equations,
in the Kerr-NUT space-time. The metric is invariant under the duality transformation involving
the exchange of mass and NUT parameters on one hand and radial and angle coordinates on the
other. We show that this invariance is also shared by the scalar and spinor perturbation equations.
Further, by the duality transformation, one can go from the Kerr to the dual Kerr solution,
and vice versa, and the same applies to the perturbation equations. In particular, it turns
out that the potential barriers felt by the incoming scalar and spinor fields are higher for
the dual Kerr than that for the Kerr. We also comment on existence of horizon and singularity.

\end{abstract}

\vskip1.0cm
\vskip2cm
\section*{I. Introduction}

For asymptotically flat axially symmetric stationary electro-vacuum space-time,
it is well known that the Kerr family is unique \cite{chandra}. On relaxing the
condition of asymptotic flatness, there occurs the NUT generalization of it, the
Kerr-NUT family \cite{nut,carter,new-dema}. It has recently been shown
\cite{dt1} that this family is also unique for the space-time admitting separable
(Hamilton-Jacobi and Klein-Gordon) equations of motion. When separability is implemented
a priori, it determines certain form of the metric functions which are the functions of radial
and angle coordinates. Then the Einstein-Maxwell equations for electro-vacuum space-time readily admit
the most general solution which turns out to be the Kerr-NUT solution \cite{new-dema} establishing
its uniqueness in a straightforward and direct manner.
When the asymptotic flatness is imposed, the NUT parameter vanishes yielding the unique
Kerr family. The Kerr family is thus included in the more general Kerr-NUT family which
is the most general family for a metric admitting separable equations of motion. It would
thus imply separability of the equations of motion for the Kerr family and it is indeed, as
is well-known, so \cite{chandra}.

The most remarkable feature of this method of obtaining the most general solution
is that it exposes the duality transformation which keeps the metric invariant. The
duality here means exchange of the mass and NUT parameters on one hand and the radial and
angular coordinates on the other \cite{dt1,dt2}. Else the Kerr-NUT solution has been known
for long \cite{new-dema}, but this invariance has thus far remained unnoticed. Like
mass is the measure of gravitational electric charge, NUT parameter
is supposed to be the measure of magnetic charge \cite{ly-no}. The Kerr-NUT solution is therefore
a truly gravitational dyon solution \cite{dt2} and it represents gravitational field
of a rotating body having both the gravitational electric (mass) and magnetic (NUT) charges.
By the duality transformation, one can obtain the space-time dual to the Kerr solution.{\footnote
{It is different from the electrogravity dual of the Kerr solution \cite{dp}.}} There exists duality
between the field of a rotating electric charge ($M$) and rotating magnetic 
charge (NUT parameter, $l$) \cite{td} and vice-versa. Though there exists pure NUT solution \cite{nut} without the
rotation parameter $a$, but for the duality transformation its presence is essential. That is,
the duality between gravitational electric and magnetic charges could be exhibited only when the body
is rotating.

The Kerr family has been studied extensively for motion of test particles as well as for the scalar,
spinor, vector and tensor perturbations of the metric. In this paper, we would like to study the scalar
and spinor perturbations, i.e. the Klein-Gordon and Dirac equations in the Kerr-NUT geometry.
As we said earlier that by a simple duality transformation one can switch over from the Kerr solution
to the dual Kerr (where $M=0$). From the properties of the Kerr metric, it is 
possible to derive the corresponding properties of the dual Kerr metric. 
For instance, the scalar and spinor perturbation equations of the Kerr 
geometry can be translated to the corresponding equations for the dual 
geometry by the duality transformation which takes the Kerr metric to the 
dual Kerr metric. However, the duality 
works only at the equation level but not at the solution level; viz the 
solution of the scalar equation in the Kerr case can not be taken over to the 
solution of the corresponding equation in the dual case by any such simple
transformations. This happens because the 
duality involves mass and NUT parameters and their presence and absence in 
the equation changes the nature of solution. In the Kerr metric there is mass 
parameter but no NUT parameter. Under the duality transformation, mass goes to 
NUT parameter and so the dual metric has NUT parameter but no mass. Hence the 
solution which is obtained with mass parameter present will have different 
character and would not go over to the one obtained with mass zero and NUT 
non-zero. In the present case, this has become more non-trivial due to the
particular selection of variable transformation which reduces the equations
into the form of wave equations whose solution is known.

The duality would have been very interesting, had it worked at the solution 
level as well. That unfortunately does not happen here 
trivially. Considering the 
Kerr solution as the field of rotating gravitational electric charge ($M$) and 
its dual as the field of rotating gravitational magnetic charge 
($l$ - the NUT parameter). Then it should be of interest to study the 
scalar and spinor perturbations of these two fields and compare their 
behaviour. This is exactly what we wish to do. We should however note that the 
dual solution is indeed an exact solution of the Einstein vacuum equation 
which is asymptotically non flat and its asymptotic flat limit is flat 
spacetime (i.e. if we impose the condition of asymptotic flatness, it reduces 
to flat spacetime). We shall probe this space-time with the scalar and spinor 
fields and compare their
motion with the corresponding motion in the Kerr geometry.
Our main aim here is to study the perturbations first in the general Kerr-NUT 
space-time and then specialize to the Kerr and the dual Kerr cases by employing the 
duality transformation. For the duality to work it is required that $a\neq0$. 
Hence we cannot go to $a=0$ limit of the pure NUT solution. As we are not
able to establish any duality relation between the solutions in the Kerr and dual
Kerr cases, we will compare the solutions graphically.

The paper is organized as follows. In \S II, we shall recall the Kerr-NUT metric and its invariance
under the duality transformation. Next two sections III and IV would respectively be devoted to the Klein-Gordon
and Dirac equations and their transformation under the duality. In \S III, we will also present the
numerical solution of scalar perturbation equation. Then in \S V, we shall present and discuss the
various numerical solutions of the Dirac spinors which would be followed by discussion of horizon and singularity in \S VI, and we end with a discussion of the main results.

\section*{II. The Kerr-NUT metric and the Duality}

Here we study the interaction of particles of spin zero and spin half
with the Kerr-NUT black hole. We are familiar
with the Klein-Gordan and Dirac equations in flat space by which one can investigate
the behaviour of spin zero and spin half particles. In the curved space-time the form of these equations are
modified. The Dirac equation was written in the Kerr geometry by Teukolsky \cite{teu1} and was separated by Chandrasekhar
\cite{chandra1} into the radial and angular parts (for a comprehensive discussion, see \cite{chandra}).
Recently, there has been a fresh investigation of the radial Dirac equation in the Schwarzschild,
Reissner-Nordstr\"om and Kerr geometries \cite{mc99,mc00,m00}.

In the linearized test field approximation,  scalar, vector and tensor
perturbations of the Kerr geometry has been studied by several authors (e.g. \cite{chandra,teu1}).
The master equations \cite{teu1} governing these linear
perturbations for integral spin (e.g., gravitational and electromagnetic)
fields were solved numerically by Press \& Teukolsky\cite{pt} and by 
Teukolsky \& Press\cite{tp}. Particularly interesting is the fact that
whereas gravitational and electromagnetic reflected wave from the black hole 
was found to be amplified for certain range of incoming frequencies, however 
Chandrasekhar\cite{chandra} predicted that no such amplification should
take place for the Dirac waves because of the very nature of the
potential experienced by the incoming fields.

The space-time metric for the Kerr-NUT geometry \cite{dt1} is given by
\begin{equation}
\label{ds2}
ds^2=-\frac{U^2}{\rho^2}\left(dt-Pd\phi\right)^2+\frac{sin^2\theta}{\rho^2}\left[(F+l^2) d\phi
-adt\right]^2+\frac{\rho^2}{U^2}dr^2+\rho^2d\theta^2,
\end{equation}
where,
\begin{equation}
F=r^2+a^2, U^2=r^2-2Mr+a^2+Q_*^2-l^2, P=asin^2\theta-2lcos\theta, \rho^2=r^2+ \lambda^2,  \lambda = l+acos\theta.
\end{equation}
Here, $l$ is the NUT parameter, $a$ and $Q_*$ are rotation and electric charge parameters of the black hole.
The most remarkable property of this metric is that it is invariant under the duality transformation, $M\leftrightarrow il,
 r\leftrightarrow i\lambda$ \cite{dt1,dt2}. That is, interchange of the gravitational electric ($M$) and magnetic ($l$) charge also
requires the corresponding interchange of the radial and angle coordinates. For the former, radial is the responding
coordinate, while for the latter, it is the angle coordinate. Further the transformation, $M\rightarrow il, r\leftrightarrow i\lambda$
will take the Kerr solution to the dual Kerr solution, and similarly vice-versa.

Now following the Newman-Penrose formalism, we introduce null tetrads $(\vec{l}, \vec{n}, \vec{m},
\vec{\bar{m}})$ to satisfy orthogonality relations, $\vec{l}{\bf .}\vec{n}=1$,
$\vec{m}{\bf .}\vec{\bar{m}}=-1$ and $\vec{l}{\bf .}\vec{m}=\vec{n}{\bf .}\vec{m}=
\vec{l}{\bf .}\vec{\bar{m}}=\vec{n}{\bf .}\vec{\bar{m}}=0$.
Thus we write the basis vectors of null tetrad in terms of elements of the Kerr-NUT geometry as
\begin{eqnarray}
\label{dnlmn}
\nonumber
l_{\mu}&=&\frac{1}{U^2}(U^2, -\rho^2, 0, -U^2 P), \hskip0.5cm
n_{\mu}=\frac{1}{2\rho^2}(U^2, \rho^2, 0, -U^2 P),\\
m_{\mu}&=&\frac{1}{\bar{\rho}\sqrt{2}}(iasin\theta, 0, -\rho^2, -i(F+l^2)sin{\theta}), \hskip0.5cm
{\bar m}_{\mu}=\frac{1}{\bar{\rho}^*\sqrt{2}}(-iasin\theta, 0, -\rho^2, i(F+l^2)sin{\theta})
\end{eqnarray}
and
\begin{eqnarray}
\label{uplmn}
\nonumber
l^{\mu}&=&\frac{1}{U^2}(F+l^2, U^2, 0, a), \hskip0.5cm
n^{\mu}=\frac{1}{2\rho^2}(F+l^2, -U^2, 0, a), \\
m^{\mu}&=&\frac{1}{\bar{\rho}\sqrt{2}}(iPcosec\theta, 0, 1, icosec{\theta}), \hskip0.5cm
{\bar m}^{\mu}=\frac{1}{\bar{\rho}^*\sqrt{2}}(-iPcosec\theta, 0, 1, -icosec{\theta}),
\end{eqnarray}
where $\bar{\rho}=r+i\lambda$ and $\bar{\rho}^*=r-i\lambda$.

In the next two sections, we study  perturbations to the Kerr-NUT (including the Kerr as well 
as the dual Kerr) space-time with the Klein-Gordan and
Dirac equations and the corresponding propagation of scalar and spinor waves respectively.
So the perturbation
can be expressed as a superposition of stationary waves with different modes
given as $e^{i(\sigma t+m\phi)}$, where
$\sigma$ is the frequency of the waves and $m$ is the azimuthal quantum number.

\section*{III. The Klein-Gordan equation}

The general equation of scalar field (wave) in curved space-time can be written as
\begin{eqnarray}
\label{kg}
\frac{1}{\sqrt{-g}}\partial_\mu\left(\sqrt{-g}g^{\mu\nu}\partial_\nu\Psi\right)-m_p^2\Psi=0
\end{eqnarray}
where $m_p$ is the mass of the
scalar field. Now substituting $\Psi=e^{i(\sigma t+m\phi)}\Phi$ in (\ref{kg}) we get
\begin{eqnarray}
\label{kg1}
\nonumber
&&\left(\frac{(F+l^2)^2}{U^2}-\frac{P^2}{sin^2\theta}\right)\sigma^2\Phi+\frac{\partial}{\partial r}(U^2
\Phi_{,r})+\frac{a(2l^2-Q_*^2+2Mr)+2lU^2cot\theta cosec\theta}{U^2}(2\sigma m)\Phi\\
&+&\left(\frac{a^2}{U^2}-\frac{1}{sin^2\theta}\right)m^2\Phi
+\frac{1}{sin\theta}\frac{\partial}{\partial \theta}(\Phi_{,\theta} sin\theta)+m_p^2\rho^2\Phi=0.
\end{eqnarray}
Further choosing $\Phi=R_0(r)S_0(\theta)$ and separating the radial and angular parts of
(\ref{kg1}) we write

\begin{eqnarray}
\label{kgth}
\left[\frac{\partial^2}{\partial \theta^2}+cot\theta\frac{\partial}{\partial \theta}-\frac{P^2\sigma^2}
{sin^2\theta}+4l\sigma mcot\theta cosec\theta+\lambda_1^2+m_p^2(l+acos\theta)^2-
\frac{m^2}{sin^2\theta}\right]S_0=0,
\end{eqnarray}
\begin{eqnarray}
\label{kgr}
\left[U^2\frac{\partial^2}{\partial r^2}+2(r-M)\frac{\partial}{\partial r}+\frac{\sigma^2}{U^2}(F
+l^2)^2+r^2m_p^2+\frac{a^2m^2}{U^2}+\frac{2a\sigma m}{U^2}(2l^2-Q_*^2+2Mr)-\lambda_1^2\right]R_0=0,
\end{eqnarray}
where $\lambda_1$ is the separation constant.

It is interesting to note that, following
\cite{dt1}, if we perform the duality transformation $M\leftrightarrow il$,
$r\leftrightarrow i\lambda$  and $R_0\leftrightarrow S_0$, and with suitable redefinition of $a$,
Eqns. (\ref{kgth}) and (\ref{kgr}) remain invariant. Further, under the duality transformation,
$M\rightarrow il$, $r\leftrightarrow i\lambda$ and $R_0\leftrightarrow S_0$, Eqns. (\ref{kgth})
and (\ref{kgr}) for the Kerr solution (with $l=0$) will go over to that of the dual Kerr solution (with $M=0$).
Similarly, the reverse duality transformation, $l\rightarrow -iM$, $r\leftrightarrow i\lambda$
and $R_0\leftrightarrow S_0$, will bring the equations back to the Kerr geometry.

Let us choose the transformation of independent variables $r$ and $\theta$ as
\begin{eqnarray}
\label{rac}
y=\frac{1}{2}log\left(\frac{1-cos\theta}{1+cos\theta}\right),\hskip0.7cm
%z=\frac{1}{2(r_+-M)}log(r-r_+)+\frac{1}{2(r_--M)}log(r-r_-),
z=\frac{r^3}{3}-(r_++r_-)\frac{r^2}{2}+r_+r_-r,
\end{eqnarray}
where $r_\pm=M\pm\sqrt{M^2+l^2-a^2-Q_*^2}$ and further
\begin{eqnarray}
\label{rz}
Z=U^2R_0.
\end{eqnarray}
Thus from (\ref{kgth}), (\ref{kgr}), (\ref{rac}) and (\ref{rz}) we get
\begin{eqnarray}
\label{ykg}
\frac{d^2S_0}{dy^2}+V_yS_0=0,
\end{eqnarray}
\begin{eqnarray}
\label{zkg}
\frac{d^2Z}{dz^2}+V_zZ=0,
\end{eqnarray}
where
\begin{eqnarray}
\label{vy}
V_y=\lambda_1^2sin^2\theta-P^2\sigma^2+4l\sigma mcos\theta+m_p^2sin^2\theta(l+acos\theta)^2-m^2,
\end{eqnarray}
\begin{eqnarray}
\label{vz}
%V_z=\sigma^2(F+l^2)^2+a^2m^2+2a\sigma m (2l^2-Q_*^2+2Mr)+r^2U^2m_p^2-\lambda_1^2U^2.
V_z=\frac{1}{U^6}\left(\sigma^2(F+l^2)^2+a^2m^2+2a\sigma m (2l^2-Q_*^2+2Mr)+r^2U^2m_p^2-
\lambda_1^2U^2+\frac{4(r-M)^2}{U^2}-2\right).
\end{eqnarray}
Here $V_y$ and $V_z$ are the effective gravitational potentials for angular and radial motion respectively of the scalar field. At
$\theta=0$ and $\pi$ and the corresponding $y=-\infty$ and $\infty$, $V_y$ reduces to $-(2l\sigma-m)^2$
and $-(2l\sigma+m)^2$ respectively. From (\ref{vy}) it is clear that $V_y$ scales with $l$ and 
$a$. Keeping the other parameters unchanged if $l$ and/or $a$ increases so does the potential. 
In the case of the radial potential, $V_z$, it falls off at large $r$ but it diverges at the 
horizon [see Eqn. (\ref{vz})]. Physically it means that the perturbation at large distance 
should not affect the space-time near the black hole. With the perturbation propagating towards 
the horizon, its effect becomes important.
Here, note that in the natural units ($G = c = h = 1$), all the parameters are dimensionless.

\begin{figure}
\vbox{
\vskip -0.5cm
\hskip 0.0cm
\centerline{
\psfig{figure=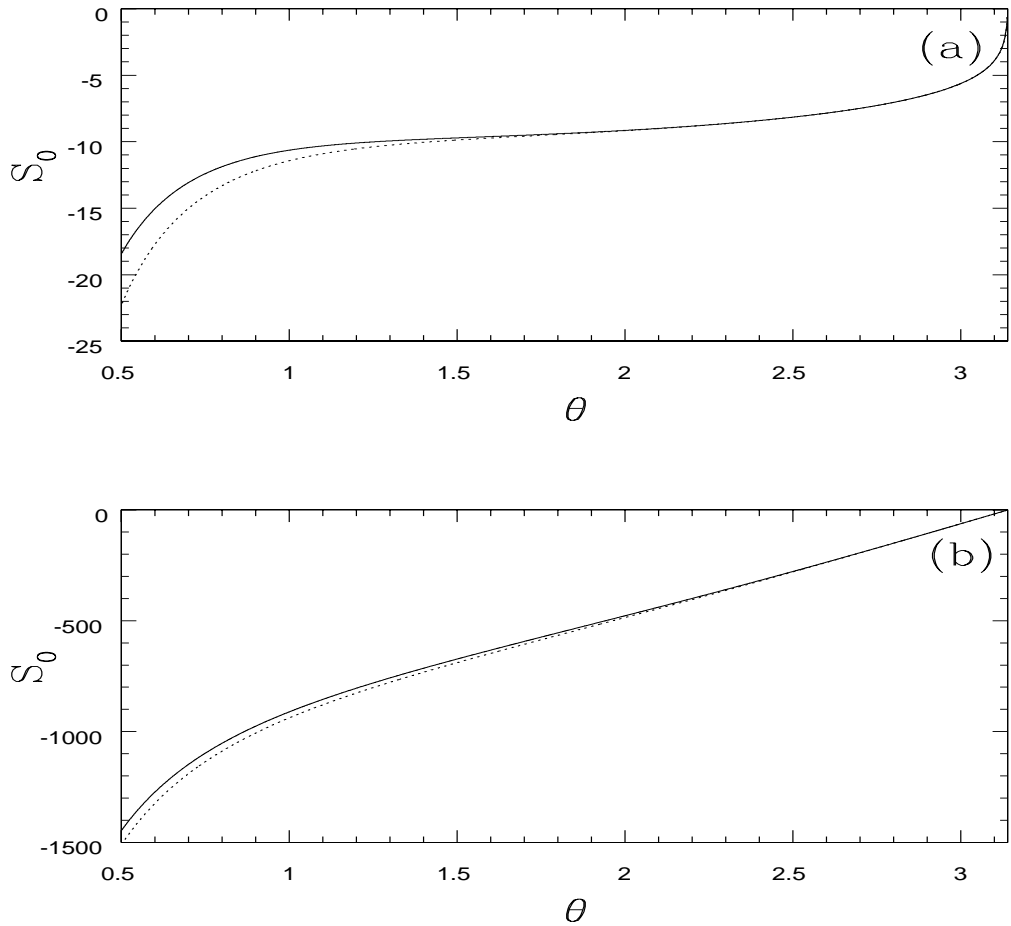,height=14truecm,width=14truecm,angle=0}}}
\vspace{-0.0cm}
%\noindent {\small {\bf Fig. 1} : 
\caption {Angular Klein-Gordan solution with $\sigma=0.4$, $M=1$ 
for (a) Kerr-NUT, $a=0.998$, $l=0.99$, (b) Kerr, $a=0.998$.  
Solid and dotted curves indicate the cases, $m_p=0.4,0.1$ respectively.
}
\label{fig1}
\end{figure}

\begin{figure}
\vbox{
\vskip -0.5cm
\hskip 0.0cm
\centerline{
\psfig{figure=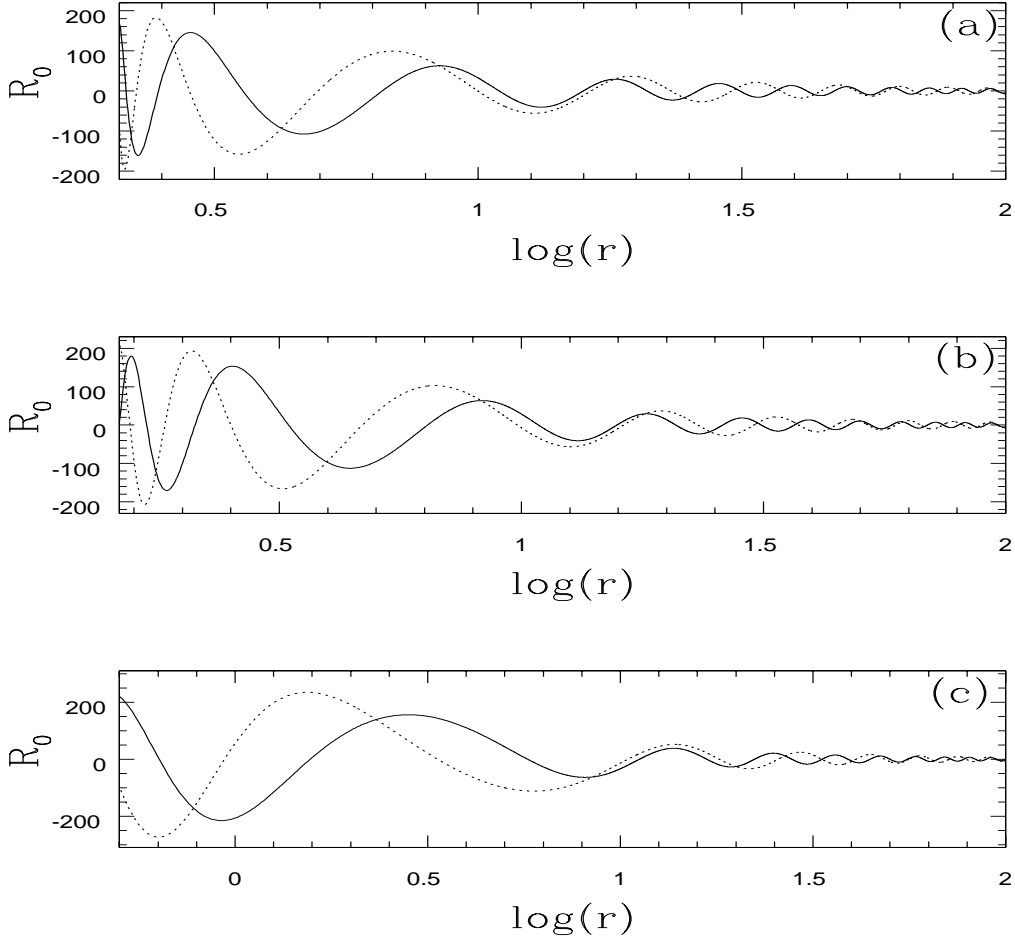,height=14truecm,width=14truecm,angle=0}}}
\vspace{-0.0cm}
%\noindent {\small {\bf Fig. 1} : 
\caption {Radial Klein-Gordan solution with $\sigma=0.4$,
for (a) Kerr-NUT, $a=0.998$, $l=0.99$, (b) Kerr, $a=0.998$, (c) dual Kerr, $a=l=0.998$.  
Solid and dotted curves indicate the cases, $m_p=0.4,0.1$ respectively. $M=1$ for (a) and (b).
}
\label{fig2}
\end{figure}

Equations (\ref{ykg}) and (\ref{zkg}) are the simple one-dimensional wave equations which we 
solve numerically and depict in Figs. 1 and 2 for the 
Kerr-NUT, Kerr and dual Kerr space-times with a certain choice of parameters.
In principle, one should first solve the angular equation with the eigenvalue, $\lambda_1$. 
Then inserting that value of $\lambda_1$ into the radial equation, one should solve
the radial equation. However, at the first instance we would like to get a qualitative feeling 
of the solution and hence we have set here $\lambda_1=1$ throughout. This is a deficiency of the 
solutions presented here which are not complete in this sense. However, they do provide some 
useful qualitative insight.
In future we plan to evaluate $\lambda_1$ and get the 
complete solution. Here our main aim is just to give an indication of form of the possible 
solution. 
We use the Runge-Kutta method with the natural boundary conditions at infinity. We know that at 
infinity
the potential barrier is either zero or constant. Therefore the wave form should be sinusoidal 
which defines the boundary condition at infinity. 
For the angular solution, first note that $V_y$ is independent of the mass $M$ and 
hence it is the same for the Kerr-NUT and the dual Kerr space-times. From the Figs. 1 and 2,
the variance of solutions as a function of the rest mass of the incoming scalar as well
as the space-time parameters, $a,l$ and $M$, is very clear. Figure \ref{fig1}
shows that the amplitude of the solution is significantly high for the Kerr case. This is because the
net gravitational effect is stronger for the Kerr case than the Kerr-NUT. 
It should be noted that the NUT and rotation parameters tend to oppose each other.
 Also the solution
shows the existence of singularity in the space-time at $\theta\rightarrow 0$. On the other hand,
Fig. \ref{fig2} indicates an extension of the inner space-time region down to  $r=0$ for the dual Kerr and
thus the wavelength of scalar waves would therefore be wider than that for the Kerr and Kerr-NUT metrics. For the radial mode, larger is the wavelength for 
less massive perturbation.

Note that the Klein-Gordon equation remains invariant under the duality transformation and the equation for
the Kerr geometry can be transformed to that for the dual Kerr geometry and vice versa, but as is
mentioned in \S 1 there is no easy
way to exploit {\it explicitly} this duality for the solutions. One of the reason behind this
is due to the certain choice of the independent variables to
cast the equation in the familiar simple form which facilitates the analysis of motion through
the behaviour of potential. The angular potential, $V_y$, depends upon the NUT parameter, $l$, 
and is free of the mass of the
black hole. The Kerr solution has only kinematic part coming from the rotation parameter, $a$.
While the radial potential, $V_z$, depends upon both $M$ and $l$. We can go to the corresponding
potential and solutions for the Kerr and the dual Kerr space-times by setting $l=0$ and $M=0$ respectively 
in the Eqns. (\ref{vy}) and (\ref{vz}).

\section*{IV. The Dirac equation}

Following \cite{m00}, derivative of spinor $P^A$ can be written as
\begin{eqnarray}
\label{ds}
D_\mu{P^A}=\partial_{\mu}P^A+iqA_\mu{P^A}+\Gamma_{\mu\nu}^A{P^\nu},
\end{eqnarray}
where $A_\mu$ and $\Gamma_{\mu\nu}^A$ are electromagnetic and gravitational
gauge fields respectively.
Thus, the Dirac equation in the Newman-Penrose formalism
can be written as
\begin{eqnarray}
\label{ds1}
\sigma_{AB'}^\mu{D_\mu}P^A+i\mu_p{\bar Q}^{C'}\epsilon_{C'B'}=0,
\end{eqnarray}
\begin{eqnarray}
\label{ds2}
\sigma_{AB'}^\mu{D_\mu}Q^A+i\mu_p{\bar P}^{C'}\epsilon_{C'B'}=0.
\end{eqnarray}
For a vector $X_i$, $\sigma_{AB'}^iX_i=X_{AB'}; A,B=0,1$; and
$2^{\frac{1}{2}}\mu_p$ is the mass of the Dirac particle.
In terms of this new basis of the Newman-Penrose formalism, the Pauli matrices can be written as
\begin{eqnarray}
\label{pauli}
\sigma_{AB'}^\mu=\frac{1}{\sqrt{2}}\left(
\begin{array}{cr} l^{\mu} & m^{\mu}\\{\bar{m}}^{\mu} & n^{\mu}\end{array}\right).
\end{eqnarray}
Following \cite{chandra} and writing
$$
P^0=F_1, P^1=F_2, {\bar Q}^{1'}=G_1, {\bar Q}^{0}=-G_2
$$
we get
\begin{eqnarray}
\label{d1}
l^{\mu}(\partial_{\mu}+iqA_\mu)F_1+{\bar m}^{\mu}(\partial_\mu+iqA_\mu)F_2+
(\epsilon-\tilde{\rho}){F_1}+(\pi-\alpha){F_2}=i{\mu_p}G_1,
\end{eqnarray}
\begin{eqnarray}
\label{d2}
m^{\mu}(\partial_{\mu}+iqA_\mu)F_1+n^{\mu}(\partial_\mu+iqA_\mu)F_2+
(\mu-\gamma){F_2}+(\beta-\tau){F_1}=i{\mu_p}G_2,
\end{eqnarray}
\begin{eqnarray}
\label{d3}
l^{\mu}(\partial_{\mu}+iqA_\mu)G_2-m^{\mu}(\partial_\mu+iqA_\mu)G_1+
({\epsilon}^*-\tilde{{\rho}}^*){G_2}-({\pi}^*-{\alpha}^*){G_1}=i{\mu_p}F_2,
\end{eqnarray}
\begin{eqnarray}
\label{d4}
n^{\mu}(\partial_{\mu}+iqA_\mu)G_1-{\bar m}^{\mu}(\partial_\mu+iqA_\mu)G_2+
({\mu}^*-{\gamma}^*){G_1}-({\beta}^*-{\tau}^*){G_2}=i{\mu_p}F_1.
\end{eqnarray}
These are the Dirac equations in the curved
space-time.

We define the spinor fields as follows:
\begin{eqnarray}
\label{sf}
f_1=e^{i({\sigma}t+m{\phi})}\bar{\rho}^*F_1,\hskip0.2cm f_2=e^{i({\sigma}t+m{\phi})}F_2,\hskip0.2cm
g_1=e^{i({\sigma}t+m{\phi})}G_1, \hskip0.2cm g_2=e^{i({\sigma}t+m{\phi})}\bar{\rho}G_2
\end{eqnarray}
and the electromagnetic vector potential, $A_{\mu}=(A_t,0,0,A_\phi)$. For the Kerr-NUT metric, it
can be expressed as
\begin{eqnarray}
\label{atph}
A_t=-\frac{rQ_{*}}{\rho^2},\hskip0.7cm A_\phi=\frac{rQ_* P}{\rho^2}.
\end{eqnarray}
We also list the spin coefficients in terms of the Kerr-NUT metric as
\begin{eqnarray}
\label{sc}
\nonumber
\tilde{\rho}&=&-\frac{1}{\bar{\rho}^*},\hskip0.5cm \beta=
\frac{cot\theta}{2\sqrt{2}{\bar{\rho}}},\hskip0.5cm \pi=
\frac{iasin\theta}{\sqrt{2}(\bar{\rho}^*)^2},\hskip0.5cm \epsilon=0,\\
\tau&=&-\frac{iasin\theta}{\rho^2\sqrt{2}},\hskip0.5cm \mu=-\frac{U^2}{2\rho^2\bar{\rho}^*},
\hskip0.5cm \gamma=\mu+\frac{r-M}{2\rho^2},\hskip0.5cm \alpha=\pi-\beta^*.
\end{eqnarray}
Now by substituting Eqns. (\ref{sf}-\ref{sc}) into Eqns. (\ref{d1}-\ref{d4}), we reduce the
Dirac equation to
\begin{eqnarray}
\label{dr1}
{\cal D}_0f_1+2^{-\frac{1}{2}}{\cal L}_{\frac{1}{2}}f_2=[i{\mu}_pr+
\mu_p(l+acos\theta)]g_1
\end{eqnarray}

\begin{eqnarray}
\label{dr2}
U^2{\cal D}^{\dagger}_{\frac{1}{2}}f_2-2^{\frac{1}{2}}
{\cal L}_{\frac{1}{2}}^{\dagger}f_1=-2[i{\mu}_pr+\mu_p(l+acos\theta)]g_2
\end{eqnarray}
\begin{eqnarray}
\label{dr3}
{\cal D}_0g_2-2^{-\frac{1}{2}}{\cal L}_{\frac{1}{2}}^{\dagger}g_1=[i{\mu}_pr-
\mu_p(l+acos\theta)]f_2
\end{eqnarray}
\begin{eqnarray}
\label{dr4}
U^2{\cal D}^{\dagger}_{\frac{1}{2}}g_1+2^{\frac{1}{2}}
{\cal L}_{\frac{1}{2}}g_2=-2[i{\mu}_pr-\mu_p(l+acos\theta)]f_1
\end{eqnarray}
where,
\begin{eqnarray}
\label{do1}
\nonumber
{\cal D}_n=\frac{d}{dr}+\frac{i}{U^2}(\sigma (F+l^2)+am+qQ_* r)
+2n\frac{r-M}{U^2}, \\
{\cal D}_n^{\dagger}=\frac{d}{dr}-\frac{i}{U^2}(\sigma (F+l^2)+am+qQ_* r)
+2n\frac{r-M}{U^2}
\end{eqnarray}
and
\begin{eqnarray}
\label{do2}
\nonumber
{\cal L}_{n}=\frac{d}{d\theta}+(\sigma P+m)cosec\theta+ncot\theta,\\
{\cal L}_n^{\dagger}=\frac{d}{d\theta}-(\sigma P+m)cosec\theta+ncot\theta.
\end{eqnarray}

Further writing $f_1(r, \theta)=R_{-\frac{1}{2}}(r)S_{-\frac{1}{2}}(\theta)$,
$f_2(r, \theta)=R_{\frac{1}{2}}(r)S_{\frac{1}{2}}(\theta)$,
$g_1(r, \theta)=R_{\frac{1}{2}}(r)S_{-\frac{1}{2}}(\theta)$,
$g_2(r, \theta)=R_{-\frac{1}{2}}(r)S_{+\frac{1}{2}}(\theta)$,
we can separate Eqns (\ref{dr1})-(\ref{dr4}) into angular and radial parts as
\begin{eqnarray}
\label{da}
%\nonumber
{\cal L}_{1/2}S_{1/2}=-({\lambda_2}-m_p(l+acos\theta))S_{-1/2},\hskip0.5cm
{\cal L}_{1/2}^{\dagger}S_{-1/2}=({\lambda_2}+m_p(l+acos\theta))S_{1/2},
\end{eqnarray}
\begin{eqnarray}
\label{dr}
%\nonumber
U{\cal D}_0R_{-1/2}=(\lambda_2+im_pr)UR_{1/2},\hskip0.5cm
U{\cal D}_0^{\dagger}UR_{1/2}=(\lambda_2-im_pr)R_{-1/2}.
\end{eqnarray}
Here, $m_p$ is the normalized rest mass of the incoming particle and $\lambda_2$
is the separation constant. Unlike the case of the scalar perturbation, the invariance of the spinor
perturbation equations under the duality transformation is non trivial, as the up and down spinor
components couple with each other. However, if we choose the duality relations between spinor components as $iS_{1/2}\leftrightarrow U^{-1/2} R_{-1/2}$ and
$S_{-1/2}\leftrightarrow U^{1/2} R_{1/2}$, Eqns.
(\ref{da}-\ref{dr}) do remain invariant when
$M\leftrightarrow il$, $r\leftrightarrow i\lambda$. As before, then we can switch over 
the Dirac equation in the Kerr space-time
to that in the dual Kerr space-time and vice versa by the duality transformation. Further the
interesting point to note is that under the duality transformation, the up-spinor transforms to
the down and vice versa, with the  multiplication/division of $U$. Next we study the separated 
angular and radial equations.

%\setcounter{equation}{7}
%\begin{eqnarray}
%\label{max7}
%\left({\cal L}_1-\frac{iasin\theta}{\bar{\rho}^*}\right)\Phi_0&=&
%\left({\cal D}_1+\frac{1}{\bar{\rho}^*}\right)\Phi_1, \nonumber \\
%\left({\cal L}_0+\frac{iasin\theta}{\bar{\rho}^*}\right)\Phi_1&=&
%\left({\cal D}_0-\frac{1}{\bar{\rho}^*}\right)\Phi_2, \\
%\left({\cal L}_1^\dagger-\frac{iasin\theta}{\bar{\rho}^*}\right)\Phi_2 &=&
%-\Delta\left({\cal D}_0^\dagger+\frac{1}{\bar{\rho}^*}\right)\Phi_1, \nonumber \\
%\left({\cal L}_0^\dagger+\frac{iasin\theta}{\bar{\rho}^*}\right)\Phi_1 &=&
%-\Delta\left({\cal D}_1^\dagger-\frac{1}{\bar{\rho}^*}\right)\Phi_0. \nonumber
%\end{eqnarray}

\subsection*{IV.A. Angular equations}

If we decouple the equations in (\ref{da}) for $S_{-{1\over 2}}$, we get
\begin{eqnarray}
\label{adec}
{\cal L}_{1\over 2}{\cal L}^\dagger_{1\over 2}S_{-{1\over 2}}+\frac{m_pasin\theta}{\lambda+
m_p(l+acos\theta)}{\cal L}^\dagger_{1\over 2}S_{-{1\over 2}}+(\lambda^2_2-m_p^2(l+acos\theta)^2)
S_{-{1\over 2}}=0.
\end{eqnarray}
Similarly one can decouple the equations in (\ref{da}) for $S_{{1\over 2}}$.
Let us choose
\begin{eqnarray}
\label{uc}
u_{\mp}=\left(\lambda_2\pm m_p l\pm\frac{m_pa}{2}\right)log(1-cos\theta)-\left(\lambda_2\pm m_p l\mp
\frac{m_pa}{2}\right)log(1+cos\theta).
\end{eqnarray}
Thus the decoupled equations for $S_{-{1\over2}}$ [i.e. Eqn. (\ref{adec})], and for 
$S_{{1\over2}}$ [not given here], can be reduced to
\begin{eqnarray}
\label{uemp}
\frac{d^2S_{\mp{1\over2}}}{du_{\mp}^2}+W_{\mp}S_{\mp{1\over2}}=0
\end{eqnarray}
where
\begin{eqnarray}
\label{upotmp}
\nonumber
W_\mp&=&\frac{sin^2\theta}{(\lambda_2\pm m_p(l+acos\theta))^2}\left[\pm(\sigma P+m)cot\theta cosec\theta
+\sigma cosec\theta (2a sin\theta cos\theta+2lsin\theta)
-\frac{cosec^2\theta}{2}+\frac{cot^2\theta}{4}\right.\\
&-&\left.(\sigma P+m)^2cosec^2\theta+\lambda_2^2-m_p^2(l+acos\theta)^2
\pm\frac{m_pasin\theta}{\lambda_2\pm m_p(l+acos\theta)}\left(\frac{cot\theta}{2}
\mp(\sigma P+m)cosec\theta\right)\right].
\end{eqnarray}

Eqn. (\ref{uemp}) is a simple one dimensional wave equation. 
%which integrates to give
%\begin{eqnarray}
%\label{uvsol}
%S_{\mp{1\over 2}}=\frac{A_\mp}{W_\mp^{1/4}} e^{i\sqrt{W_\mp}u_\mp}+\frac{B_\mp}{W_\mp^{1/4}} e^{-i\sqrt{W_\mp}u_\mp}.
%\end{eqnarray}
Here again the angular potential depends upon $l$ and is free off $M$. 

The above gravitational potentials are related to the physical parameters and variable
in a rather complicated manner compared to the case of scalar perturbation. 
For both $\theta\rightarrow 0$ and $\pi$, angular potential 
barriers reduce to zero ($W_\pm\rightarrow 0$). 
We can study the potential barriers felt by the incoming spinors with different
masses and frequencies. In principle one can choose any mass and frequency of the field but to
bring a significant interaction with the black hole we choose the mass and frequency
in such a manner that the Compton wavelength of the incoming
fields becomes same order as the radius of the outer horizon to the black hole. Also the frequency
of the perturbation should be same order as the inverse of the light crossing time to the
radius of the black hole. Thus we choose
\begin{eqnarray}
\label{masssig}
\sigma\sim m_p\sim r_+^{-1}.
\end{eqnarray}

\begin{figure}
\vbox{
\vskip -0.5cm
\hskip 0.0cm
\centerline{
\psfig{figure=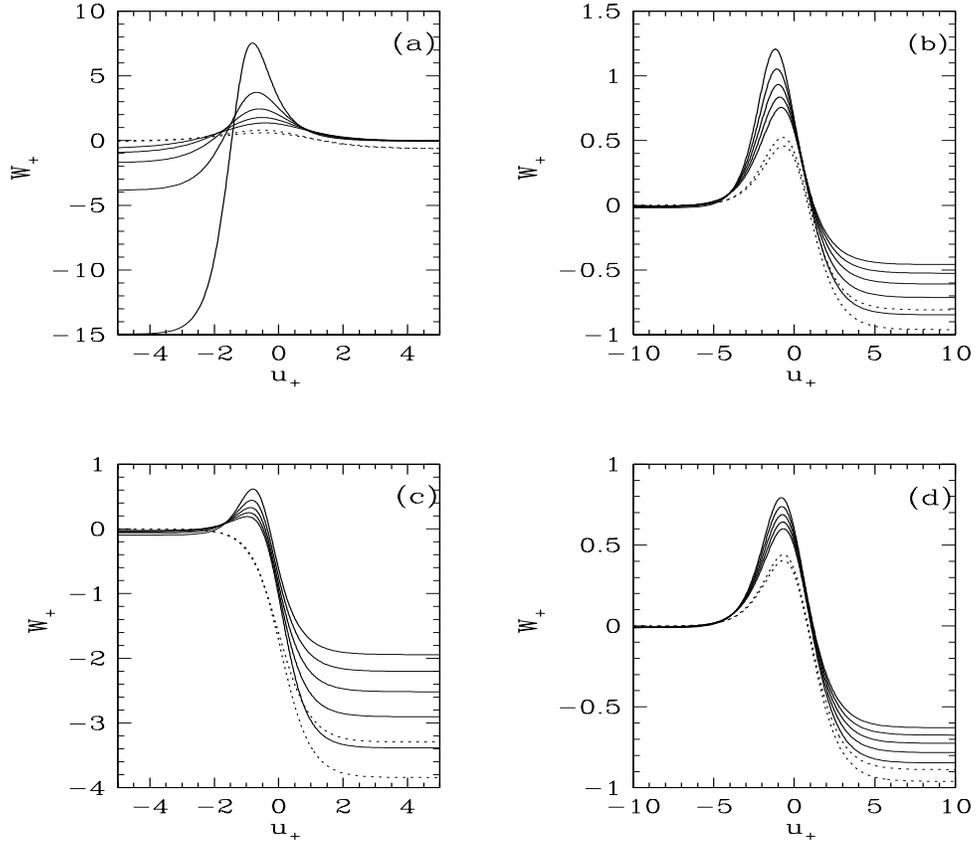,height=14truecm,width=14truecm,angle=0}}}
\vspace{-0.0cm}
\caption{ Variation of $W_+$ as a function of $u_+$, $M=1$
when (a) $a=0.998$, $l=0.99$, (b) $a=0.998$, $l=0.1$, (c) $a=l=0.5$, (d) $a=0.5$, $l=0.1$.
Solid curves indicate the cases, $\sigma=0.4$; $m_p=0.4,0.3,0.2,0.1,0$ from top to bottom and dotted curves
indicate $\sigma=0.1$; $m_p=0.1,0$ from top to bottom.
}
\label{fig3}
\end{figure}

\begin{figure}
\vbox{
\vskip -0.5cm
\hskip 0.0cm
\centerline{
\psfig{figure=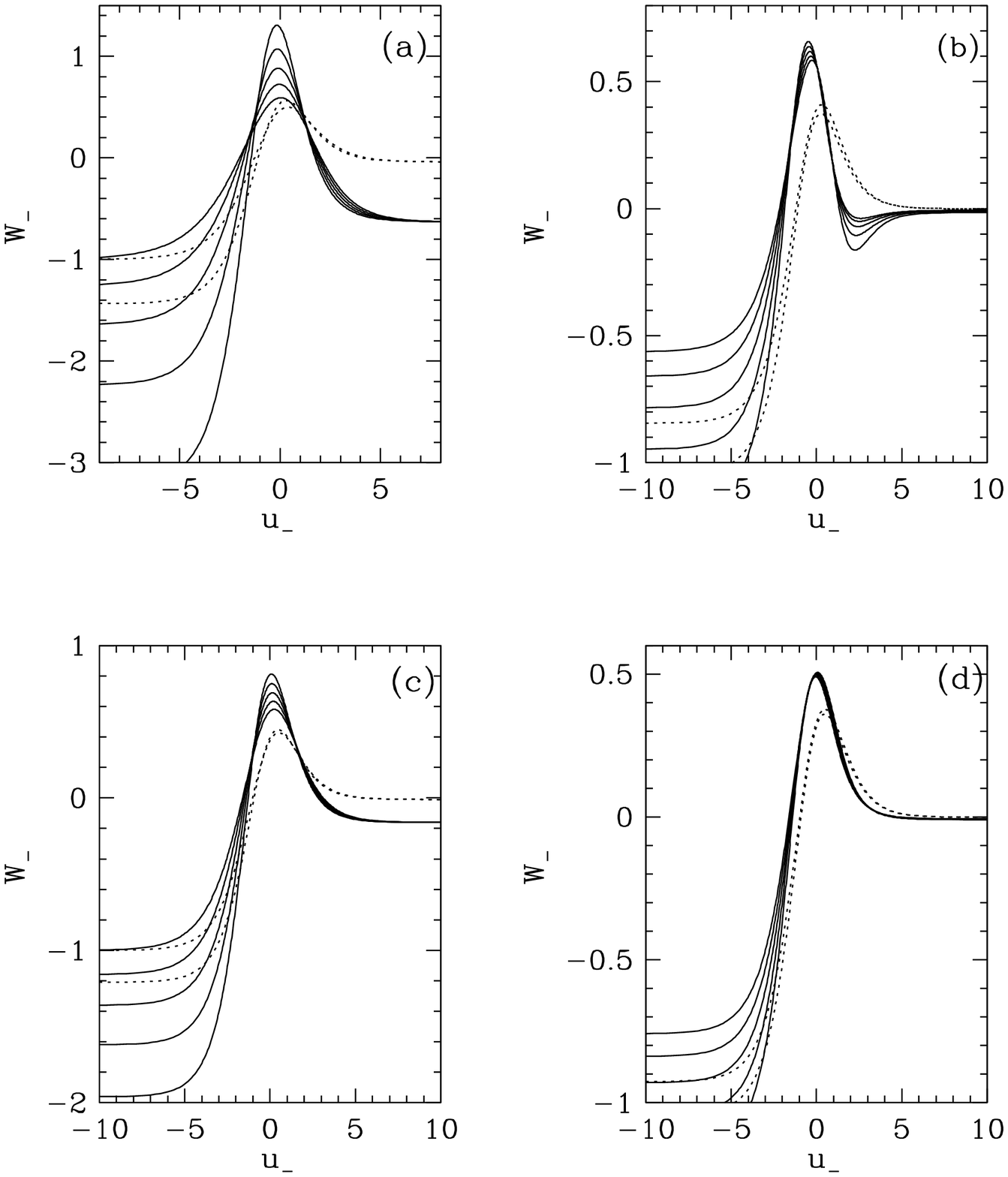,height=14truecm,width=14truecm,angle=0}}}
\vspace{-0.0cm}
\caption{ Variation of $W_-$ as a function of $u_-$, $M=1$
when (a) $a=0.998$, $l=0.99$, (b) $a=0.998$, $l=0.1$, (c) $a=l=0.5$, (d) $a=0.5$, $l=0.1$.
Solid curves indicate the cases, $\sigma=0.4$; $m_p=0.4,0.3,0.2,0.1,0$ from top to bottom and dotted curves
indicate $\sigma=0.1$; $m_p=0.1,0$ from top to bottom.
}
\label{fig4}
\end{figure}

Figures \ref{fig3} and \ref{fig4} indicate the behaviour of potentials felt by the spinors in the angular 
direction in the transformed coordinate system. For a particular $\sigma$, as $m_p$ decreases the
coupling strength between the spinor and space-time curvature reduces then the peak of the barrier
goes down. Similarly for the same $m_p$ if $\sigma$ reduces the peak goes down. The interesting 
feature that emerges out is that the height of the barrier is quite sensitive to the spin orientation 
(up/down) of the incoming spinor field. This is in accordance with the expectation that a 
spinning black hole (with a particular spin orientation) can distinguish 
the up and down spin of the perturbation separately. Further a spinor feels higher potential 
barrier when it is aligned with
the black hole. We also note the mirror inversion symmetry between $W_+$ and 
$W_-$.

\subsection*{IV.B. Radial equations}

Decoupling the equations in Eqn. (\ref{dr}) for $R_{-{1\over 2}}$ we write
\begin{eqnarray}
\label{er}
U^2{\cal D}^\dagger_{1\over 2}{\cal D}_0R_{-{1\over 2}}-\frac{im_p U^2}{\lambda_2+
im_p r}{\cal D}_0R_{-{1\over 2}}-(\lambda_2^2+m_p^2r^2)R_{-{1\over 2}}=0.
\end{eqnarray}
However, following the earlier works \cite{chandra,mc00}, we cast this into the form,
\begin{eqnarray}
\label{erd}
\frac{d^2Z_\pm}{d\hat{r}_*^2}+\left(\sigma^2-V_\pm\right)Z_\pm=0
\end{eqnarray}
which is easier to attack. Here we define
\begin{eqnarray}
\label{z}
Z_\pm=UR_{{1\over 2}}e^{i\Theta/2}\pm R_{-{1\over 2}}e^{-i\Theta/2},\hskip0.5cm
\Theta=tan^{-1}\left(\frac{m_p r}{\lambda}\right),
\end{eqnarray}
\begin{eqnarray}
\label{rsh}
\nonumber
\hat{r}_*&=&r+\frac{2Mr_+-Q_*^2+2l^2+\frac{am+qQ_*r_+}{\sigma}}{r_+-r_-}log\left(\frac{r}{r_+}-1\right)
-\frac{2Mr_--Q_*^2+2l^2+\frac{am+qQ_*r_-}{\sigma}}{r_+-r_-}log\left(\frac{r}{r_-}-1\right)\\
&+&\frac{1}{2\sigma}tan^{-1}\left(\frac{m_p r}{\lambda}\right),
\end{eqnarray}
\begin{eqnarray}
\label{v}
\nonumber
V_\pm&=&{{U(\lambda_2^{2} + m_p^{2} r^{2})^{3/2}} \over {[ \omega^{2}(\lambda_2^{2} + m_p^{2}
r^{2}) + \frac{\lambda_2 m_p U^2}{2 \sigma}]^{2}}}[U(\lambda_2^{2} + m_p^{2} r^{2})^{3/2} \pm
 \{(r-M)(\lambda_2^{2} + m_p^{2} r^{2}) + 3m_p^{2} r U^2\}]\\
& \mp& {{U^{3}(\lambda_2^{2} + m_p^{2} r^{2})^{5/2}} \over {[ \omega^{2}(\lambda_2^{2} + m_p^{2}
 r^{2}) + \frac{\lambda_2 m_p U^2}{2 \sigma}]^{3}}}\left[\left(2r+\frac{qQ_*}{\sigma}\right)(\lambda_2^{2} +
 m_p^{2} r^{2}) + 2 m_p^{2} \omega^{2} r + \frac{\lambda_2 m_p (r-M)}{\sigma}\right],
\end{eqnarray}
where
\begin{eqnarray}
\omega^2=r^2+\alpha^2, ~\, \alpha^2 = a^2+l^2+\frac{am}{\sigma}+\frac{qQ_*r}{\sigma}.
\end{eqnarray}
%Therefore, Eqn. (\ref{erd}) solves to give
%\begin{eqnarray}
%\label{zsol}
%Z_\pm=\frac{C_\pm}{\sqrt{k_\pm}} e^{ik_\pm\hat{r}_*}+\frac{D_\pm}{\sqrt{k_\pm}} e^{-ik_\pm\hat{r}_*},
%\end{eqnarray}
%where $k_\pm=\sqrt{\sigma^2-V_\pm}$.

From Eqn. (\ref{rsh}), it is clear that for
\begin{eqnarray}
\label{sigs}
\sigma=\sigma_s=-\frac{qQ_* r_++am}{2(Mr_++l^2)-Q_*^2},
\end{eqnarray}
coefficient of $log\left(\frac{r}{r_+}-1\right)$ vanishes and thus for $\sigma\leq\sigma_s$, the 
relation  
$\hat{r}_*-r$ becomes multivalued. From Eqn. (\ref{v}) it is clear that the potential, $V_\pm$, diverges at $r=|\alpha|$.
This regime where $V_\pm$ diverges, is well known as super-radiance.
For $\sigma >\sigma_s$, $|\alpha | <r_+$, and the wave can not reach at $r=|\alpha |$ and hence there is 
no question of super-radiance. For $\sigma=\sigma_s$, $|\alpha |=r_+$, and  the potential 
[Eqn. (\ref{v})] diverges exactly at
the event horizon. For $\sigma>\sigma_s$, $\hat{r}_*$ runs from $-\infty$ to $\infty$ as $r$ varies
from $r_+$ to $\infty$; but for $\sigma\leq \sigma_s$, at both $r\rightarrow r_+$ and $r\rightarrow \infty$,
$\hat{r}_*\rightarrow\infty$. The turning point of the $\hat{r}_*-r$ relation occurs at 
$r=|\alpha|$. Thus the analogue of ergo-sphere is defined by the region $r_+\leq r\leq |\alpha|$. 
However, the spinor particles do not experience super-radiance and hence no energy can be 
extracted by the spinor perturbation (e.g. \cite{chandra,mc00,wd}). However the result is 
non null for the electromagnetic
perturbation and energy is possible to extract from the black hole.

It is clear from Eqn. (\ref{v}) that, like angular potential barrier, here also the gravitational potentials are 
related to the physical parameters and variable in a complicated manner. Close to the
horizon, the potential barrier, $V_\pm$, reduces to zero but at a
large distance $V_\pm\rightarrow m_p^2$. Thus from Eqn. (\ref{erd}), it
appears that $\sigma$ has to be greater than or equal to $m_p$, otherwise spinors can not
enter into the gravitational field. 
    
\begin{figure}
\vbox{
\vskip -0.5cm
\hskip 0.0cm
\centerline{
\psfig{figure=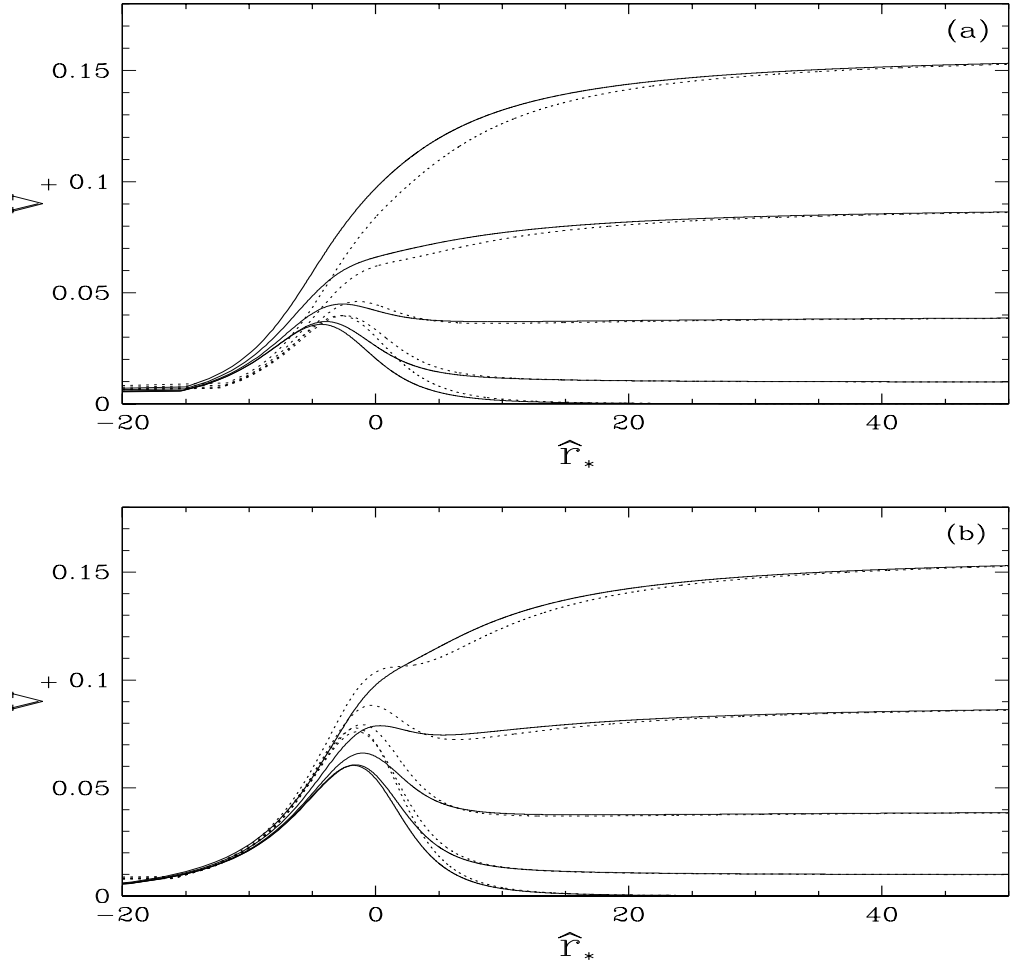,height=14truecm,width=14truecm,angle=0}}}
\vspace{-0.0cm}
\caption{ Variation of $V_+$ as a function of $\hat{r}_*$ for $\sigma>\sigma_s$,
when (a) $a=0.998$, $l=0.99$, (b) $a=0.998$, $l=0.1$. Solid curves are for $\sigma=0.7$ and dotted
curves for $\sigma=0.4$. From the top to bottom $m_p$ varies as $0.4,0.3,0.2,0.1,0$, and $M=1$, $m=-1/2$, $Q_*=0$.
}
\label{fig5}
\end{figure}

\begin{figure}
\vbox{
\vskip -0.5cm
\hskip 0.0cm
\centerline{
\psfig{figure=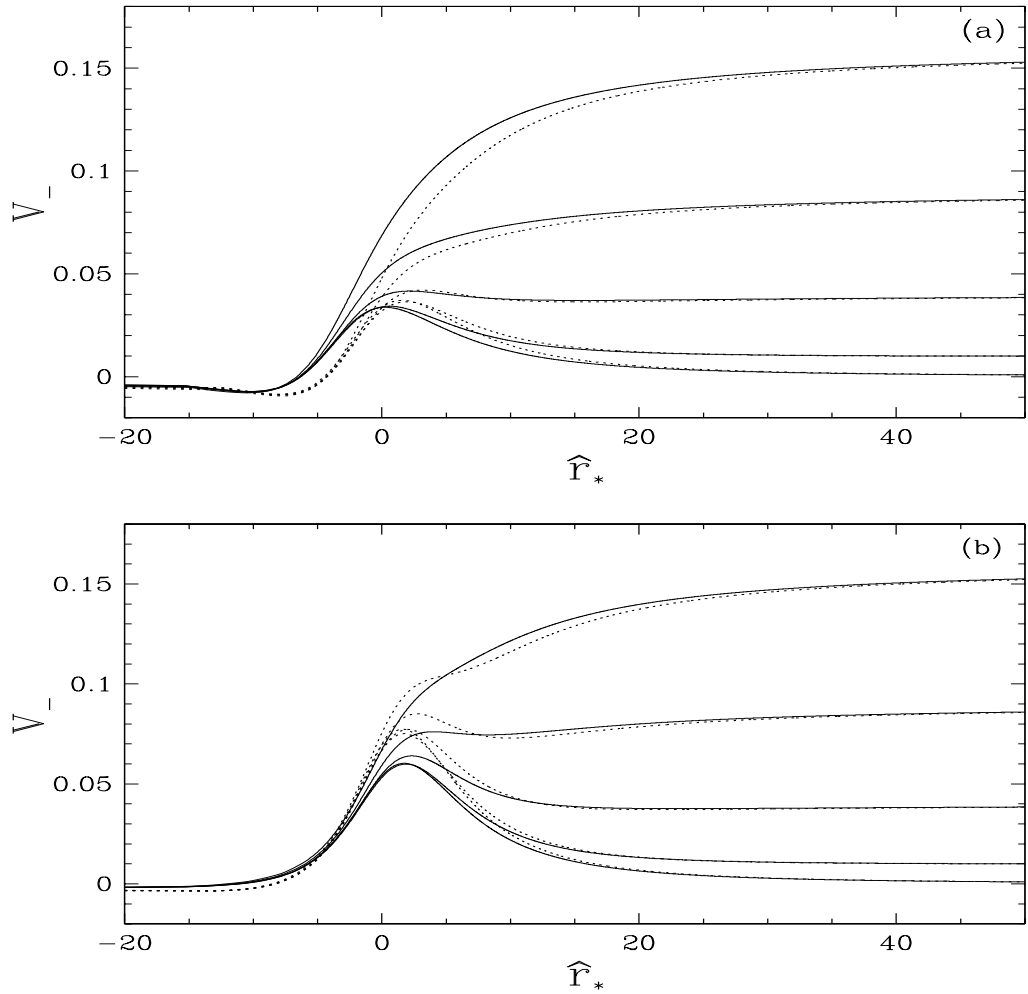,height=14truecm,width=14truecm,angle=0}}}
\vspace{-0.0cm}
\caption{ Variation of $V_-$ as a function of $\hat{r}_*$ for $\sigma>\sigma_s$,
when (a) $a=0.998$, $l=0.99$, (b) $a=0.998$, $l=0.1$. Solid curves are for $\sigma=0.7$ and dotted
curves for $\sigma=0.4$. From the top to bottom $m_p$ runs through $0.4,0.3,0.2,0.1,0$, and $M=1$, $m=-1/2$, $Q_*=0$.
}
\label{fig6}
\end{figure}

Figures \ref{fig5} and \ref{fig6} show the behaviour of potential barriers ($V_\pm$) for 
different choices of the frequency and
mass of the spinor. Here also we choose $\sigma\sim m_p\sim r_+^{-1}$,
to consider a significant interaction between the incoming spinor and the black hole.
For a particular frequency, as the mass of the spinor decreases the asymptotic
height of the barrier decreases. 
     
\begin{figure}
\vbox{
\vskip -0.5cm
\hskip 0.0cm
\centerline{
\psfig{figure=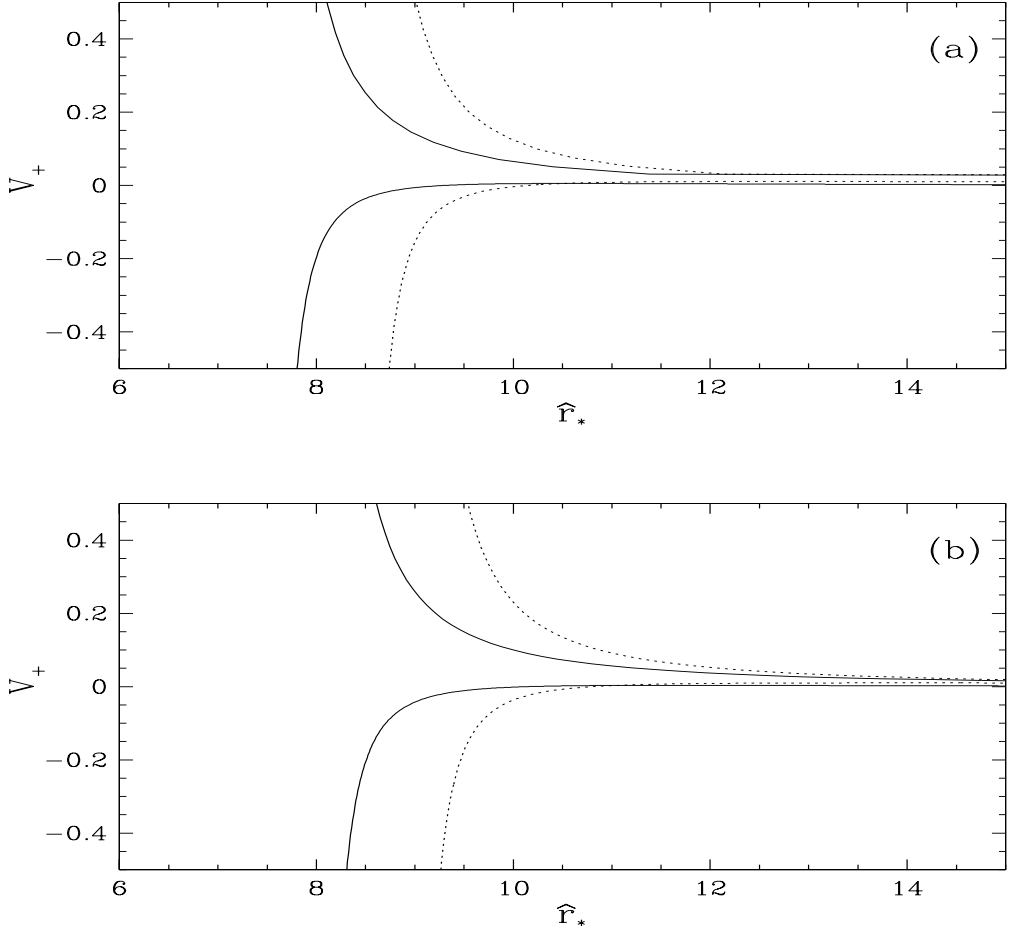,height=14truecm,width=14truecm,angle=0}}}
\vspace{-0.0cm}
\caption{ Variation of $V_+$ as a function of $\hat{r}_*$ for $\sigma\leq\sigma_s$
($\sigma=0.1$), when (a) $a=0.998$, $l=0.5$, (b) $a=0.998$, $l=0.1$. Solid and dotted curves indicate  $m_p=0, 0.1$ respectively with $M=1$, $m=-1/2$, $Q_*=0$
}
\label{fig7}
\end{figure}

In Fig. \ref{fig7}, the behaviour of potentials ($V_+$) in super-radiance regime
is shown. Clearly, at a certain $r$ the potential diverges. For a particular $\sigma$, if $m_p$
increases, the singular point shifts outward. The behaviour of corresponding $V_-$ is similar.

\subsection*{IV.C. Comparison of Kerr with dual Kerr}

For the dual Kerr solution $M=0$, the corresponding $\hat{r}_*$ would read as
\begin{eqnarray}
\label{rshv}
\hat{r}_*=r-\frac{\alpha^2}{r}+\frac{1}{2\sigma}tan^{-1}\left(\frac{m_p r}{\lambda}\right)
\end{eqnarray}
for $a=l$, and
\begin{eqnarray}
\label{rshv}
\hat{r}_*=r+\frac{2l^2+\frac{a\sigma}{m}}{\sqrt{a^2-l^2}}tan^{-1}\frac{r}{\sqrt{a^2-l^2}}
+\frac{1}{2\sigma}tan^{-1}\left(\frac{m_p r}{\lambda}\right)
\end{eqnarray}
for $a>l$. The radial potential would depend upon $a$ and $l$ only, keeping unchanged
the form of Eqn. (\ref{v}). 

\begin{figure}
\vbox{
\vskip -0.5cm
\hskip 0.0cm
\centerline{
\psfig{figure=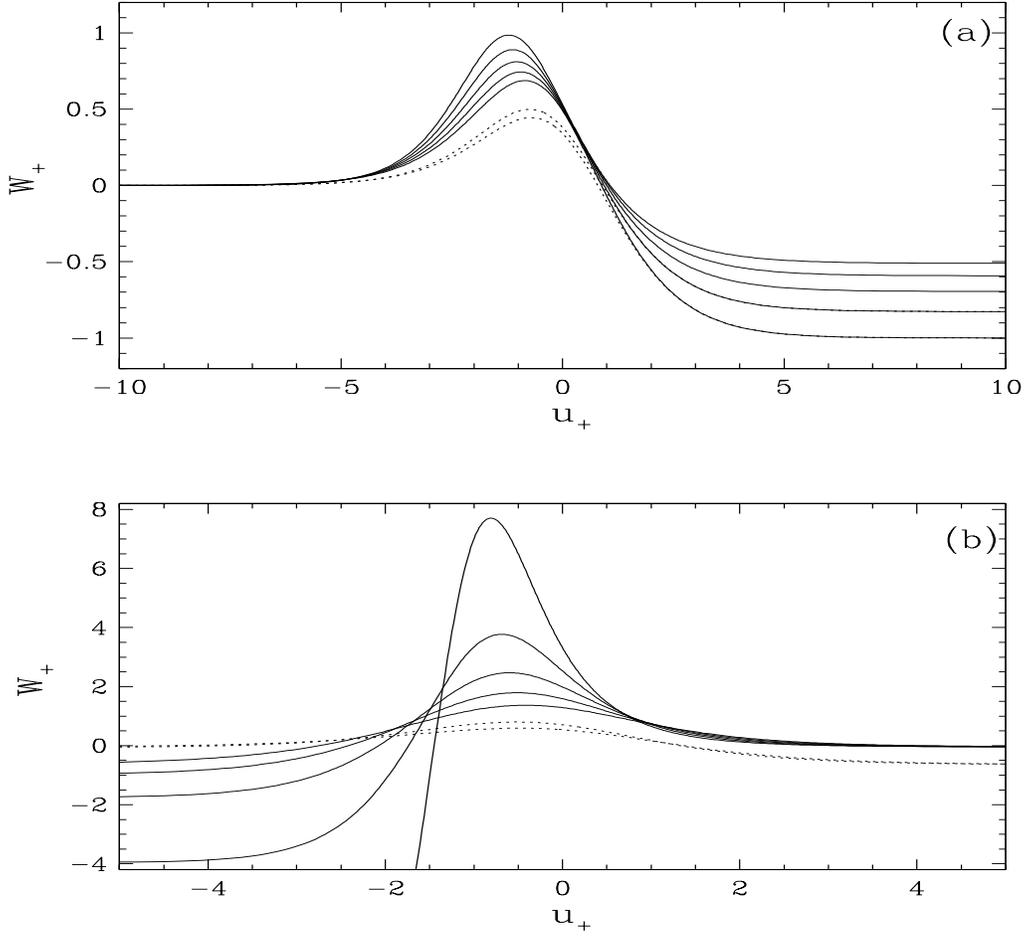,height=14truecm,width=14truecm,angle=0}}}
\vspace{-0.0cm}
\caption{ Variation of $W_+$  for the dual Kerr ($M=0$) case as a function of $u_+$ for 
(a) $a=0.998$, $l=0$, $M=1$ (Kerr) (b) $a=l=0.998$.
From top to bottom, solid curves are for $\sigma=0.4$; $m_p=0.4,0.3,0.2,0.1,0$ while dotted are for $\sigma=0.1$; $m_p=0.1,0$, and $m=-1/2$, $Q_*=0$.
}
\label{fig8}
\end{figure}

\begin{figure}
\vbox{
\vskip -0.5cm
\hskip 0.0cm
\centerline{
\psfig{figure=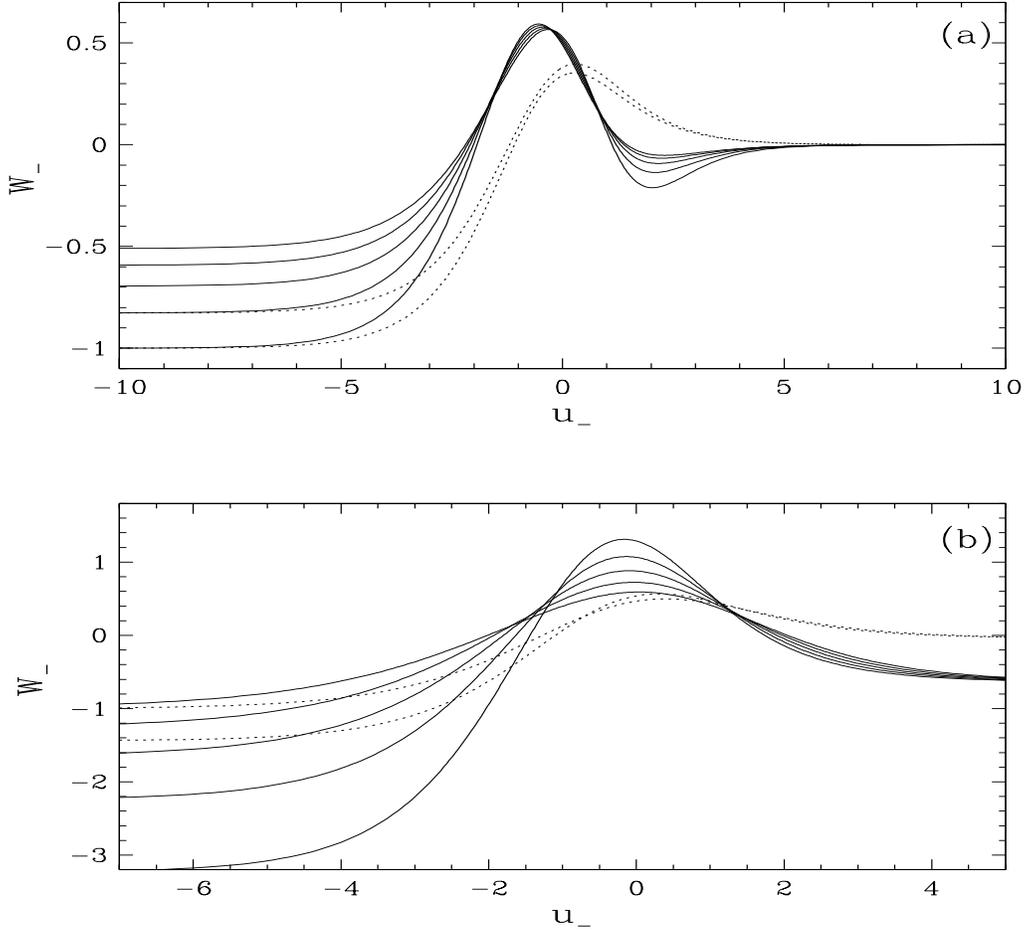,height=14truecm,width=14truecm,angle=0}}}
\vspace{-0.0cm}
\caption{ Potential $W_-$ is plotted against $u_-$ for  
(a) $a=0.998$, $l=0$, $M=1$ (Kerr) (b) $a=l=0.998$, $M=0$ (dual Kerr).
From top to bottom, solid curves are for $\sigma=0.4$; $m_p=0.4,0.3,0.2,0.1,0$ and dotted are for $\sigma=0.1$; $m_p=0.1,0$, and $m=-1/2$, $Q_*=0$.
}
\label{fig9}
\end{figure}

Figures \ref{fig8} and \ref{fig9} show the variation of angular potential in the Kerr and dual of Kerr metrics. It is
seen that the peak of the barrier (and the maximum height of it) in the dual Kerr for a particular set of
$\sigma$ and $m_p$ is higher than that in the Kerr case. However $W_\pm$ in the dual Kerr is 
the same as for the Kerr-NUT
for the angular motion because of its $M$ independence. As before, here again $W_+$ is 
higher than $W_-$ because of the opposite spin orientation of the perturbation, 
particularly in the higher frequency regime.
	  
\begin{figure}
\vbox{
\vskip -0.5cm
\hskip 0.0cm
\centerline{
\psfig{figure=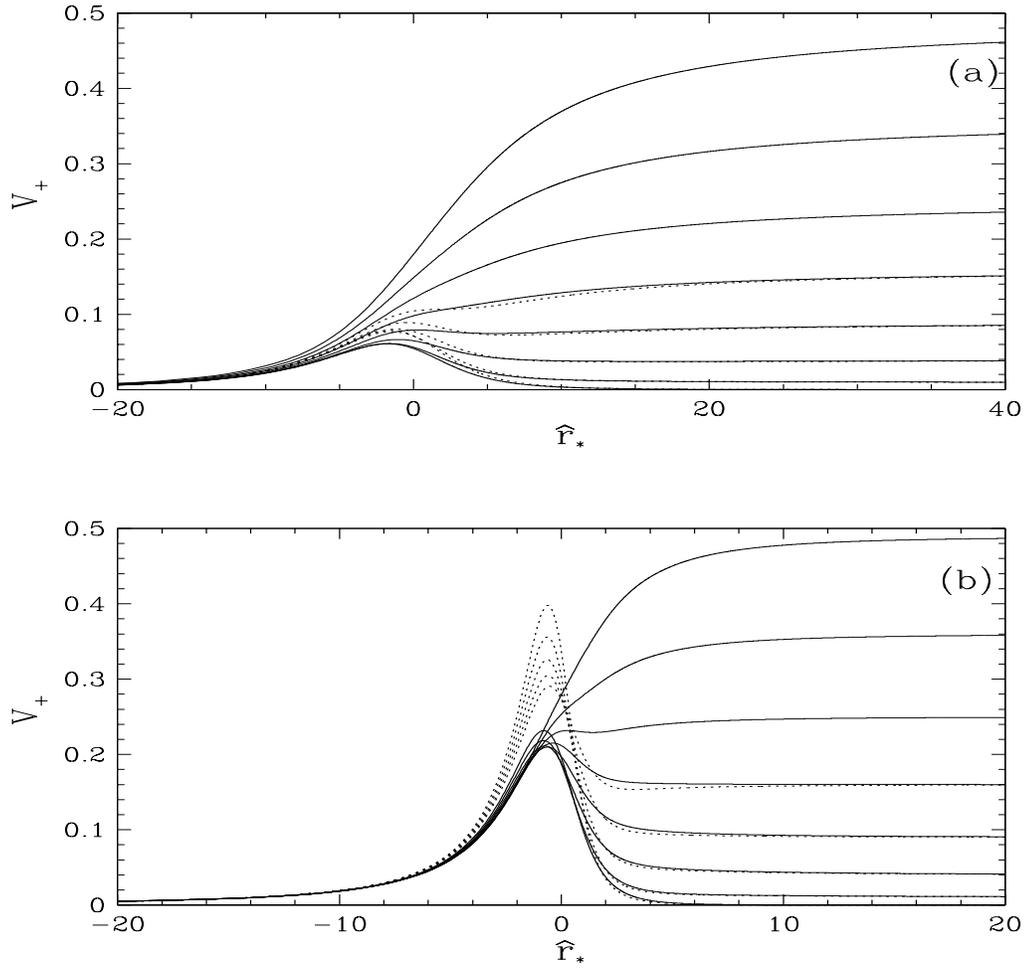,height=14truecm,width=14truecm,angle=0}}}
\vspace{-0.0cm}
\caption{ Variation of $V_+$ as a function of $\hat{r}_*$, 
when (a) $a=0.998$, $l=0$, $M=1$ (Kerr) (b) $a=l=0.998$, $M=0$ (dual Kerr).
Solid curves are from top to bottom for $\sigma=0.7$; $m_p=0.7,0.6,0.5,0.4,0.3,0.2,0.1,0$ and dotted
are for $\sigma=0.4$; $m_p=0.4,0.3,0.2,0.1,0$, and $m=-1/2$, $Q_*=0$.
}
\label{fig10}
\end{figure}

\begin{figure}
\vbox{
\vskip -0.5cm
\hskip 0.0cm
\centerline{
\psfig{figure=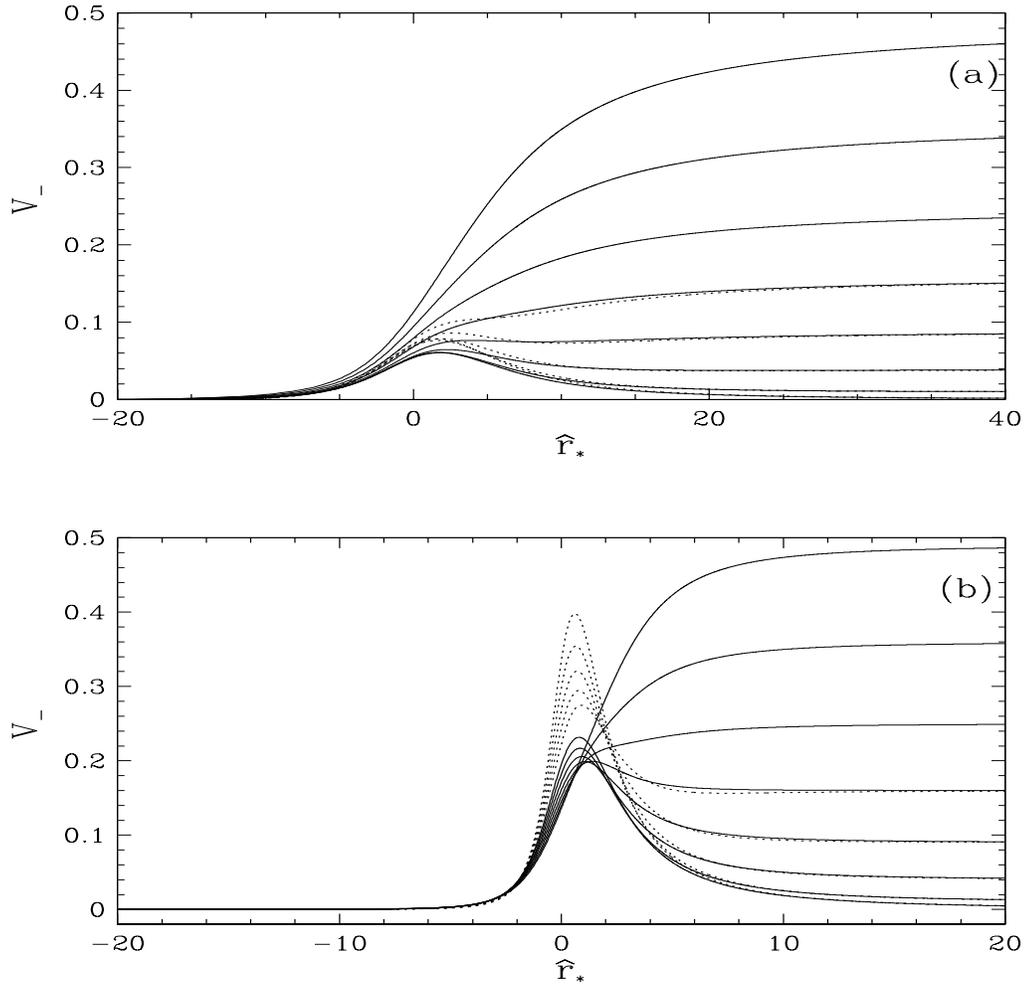,height=14truecm,width=14truecm,angle=0}}}
\vspace{-0.0cm}
\caption{ Variation of $V_-$ as a function of $\hat{r}_*$, 
when (a) $a=0.998$, $l=0$, $M=1$ (Kerr) (b) $a=l=0.998$, $M=0$ (dual Kerr).
Solid curves are from top to bottom for $\sigma=0.7$; $m_p=0.7,0.6,0.5,0.4,0.3,0.2,0.1,0$ and dotted 
are for $\sigma=0.4$; $m_p=0.4,0.3,0.2,0.1,0$, and $m=-1/2$, $Q_*=0$.
}
\label{fig11}
\end{figure}

In Figs. \ref{fig10} and \ref{fig11}, we compare the radial gravitational potential in the 
Kerr and dual Kerr cases. It is again
seen that the peak of the barrier (and the maximum height of it) in the dual Kerr for a particular set of
$\sigma$ and $m_p$ is higher than that in the Kerr. In the dual Kerr, for a particular $m_p$, if
$\sigma$ decreases, the peak of the barrier distinctly increases.

Therefore, from the above discussion we now have a clear picture of the behaviour of 
gravitational potential barriers felt by the spinors in the Kerr-NUT, Kerr
and dual Kerr space-times. We will next study their solutions.

\section*{V. Solution for spinor perturbation}

Here we like to solve Eqns. (\ref{uemp}) and (\ref{erd}) for the various sets of the physical
parameters. In view of the stationary and axisymmetric nature of the background space-time, 
it is natural to write the perturbation as superposition of waves of different modes, 
$exp[i(\sigma t+m\phi)]$. We have thus to solve the radial and angular equations. 
These equations are however not solvable analytically and hence we are forced to seek for
numerical solutions. We shall employ the well known Runge-Kutta method with the boundary
conditions at infinity. With the aim of getting the qualitative feeling of the solution, 
we shall again set $\lambda_2=1$, which should in principle be evaluated in an exact manner. 
The complete solution has been obtained for the Kerr geometry \cite{chak}. We shall however 
defer this for a future consideration for the Kerr-NUT geometry.

 The numerical solutions are shown in the following figures. The boundary condition at infinity 
is fixed by demanding sinusoidal wave with the wave number, $k_\pm=\sqrt{\sigma^2-V_\pm}$.

\begin{figure}
\vbox{
\vskip -0.5cm
\hskip 0.0cm
\centerline{
\psfig{figure=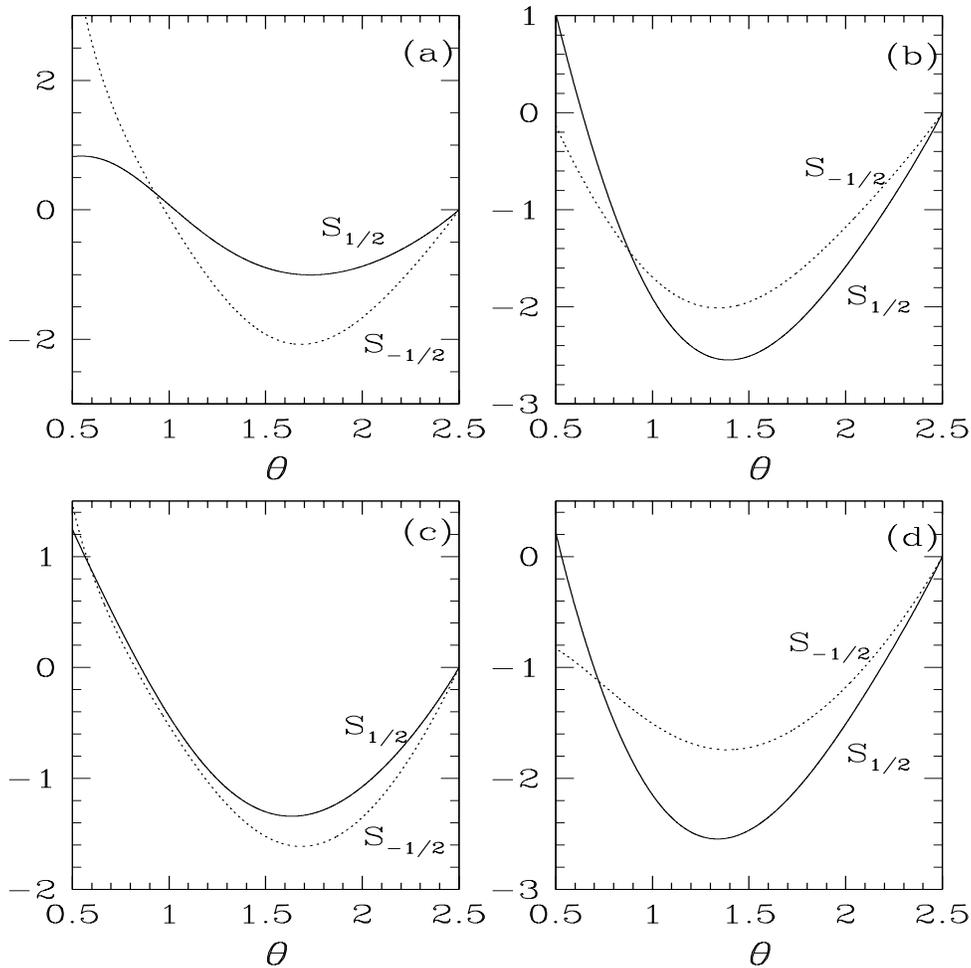,height=14truecm,width=14truecm,angle=0}}}
\vspace{-0.0cm}
\caption{ Angular spin-up and spin-down solutions with $\sigma=m_p=0.4$,
for (a) Kerr-NUT, $a=0.998$, $l=0.99$ (b) Kerr, $a=0.998$, 
(c) Kerr-NUT, $a=l=0.5$, (d) Kerr, $a=0.5$.
Solid and dotted curves indicate the up and down spinors respectively as marked in the
figure too. Other parameters are $m=-1/2$, $Q_*=0$.
}
\label{fig12}
\end{figure}

Figure \ref{fig12} represents the behaviour of spinors in the angular direction for the Kerr-NUT and
Kerr geometry. Recall from Figs. \ref{fig3} and \ref{fig4} that potential (which is also the square 
of wave number according to our notation) is positive only in the vicinity of $\theta=\pi/2$, 
then (\ref{uemp}) can admit a harmonic oscillator like solution. We will essentially be interested 
in this region only because the regions of negative $W_\pm$ can only give exponentially
increasing and/or decreasing solutions. The solutions clearly bring out the 
role played by the NUT parameter. 

\begin{figure}
\vbox{
\vskip -0.5cm
\hskip 0.0cm
\centerline{
\psfig{figure=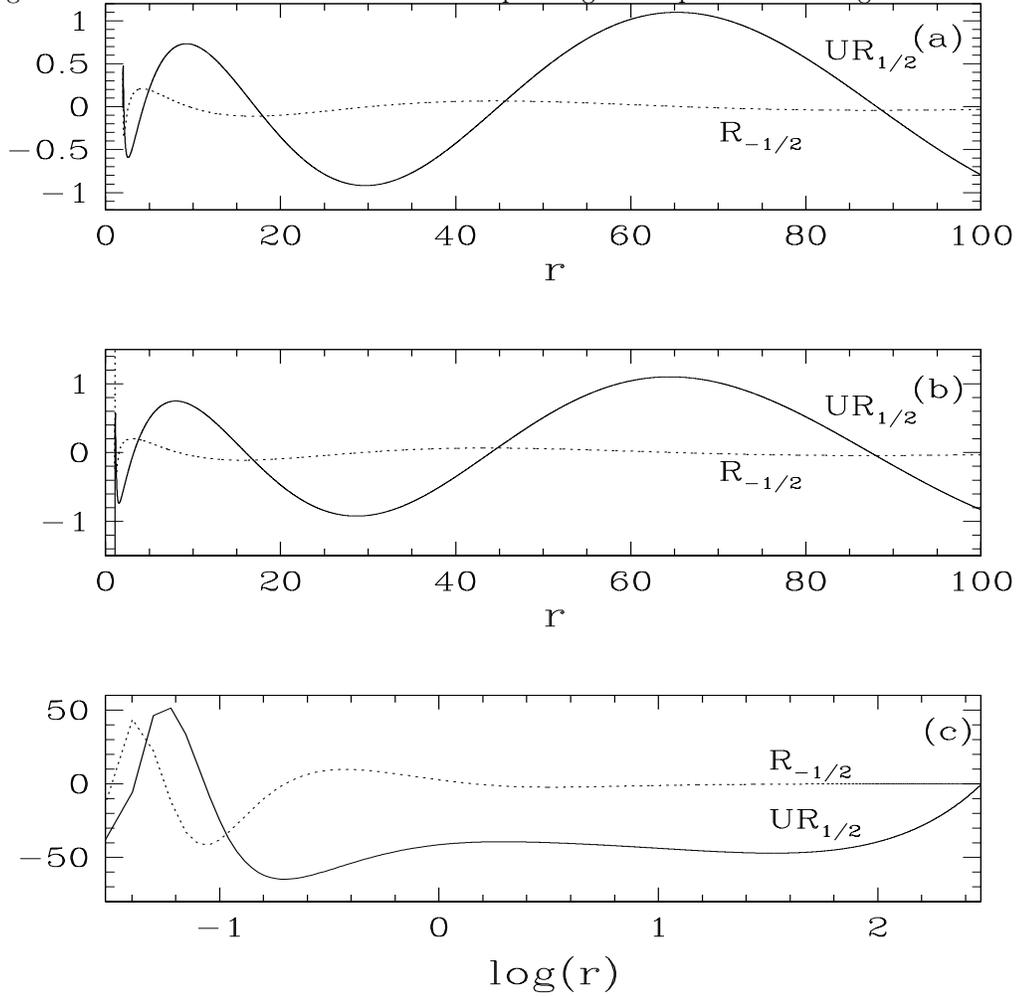,height=14truecm,width=14truecm,angle=0}}}
\vspace{-0.0cm}
\caption{ Radial spin-up and spin-down solutions with $\sigma=m_p=0.4$,
for (a) Kerr-NUT, $a=0.998$, $l=0.99$ (b) Kerr, $a=0.998$, 
(c) dual Kerr, $a=l=0.998$.
Solid and dotted curves indicate the up and down spinors respectively as marked in the
figure too. Other parameters are $m=-1/2$, $Q_*=0$; for (a) and (b) $M=1$.
}
\label{fig13}
\end{figure}

\begin{figure}
\vbox{
\vskip -0.5cm
\hskip 0.0cm
\centerline{
\psfig{figure=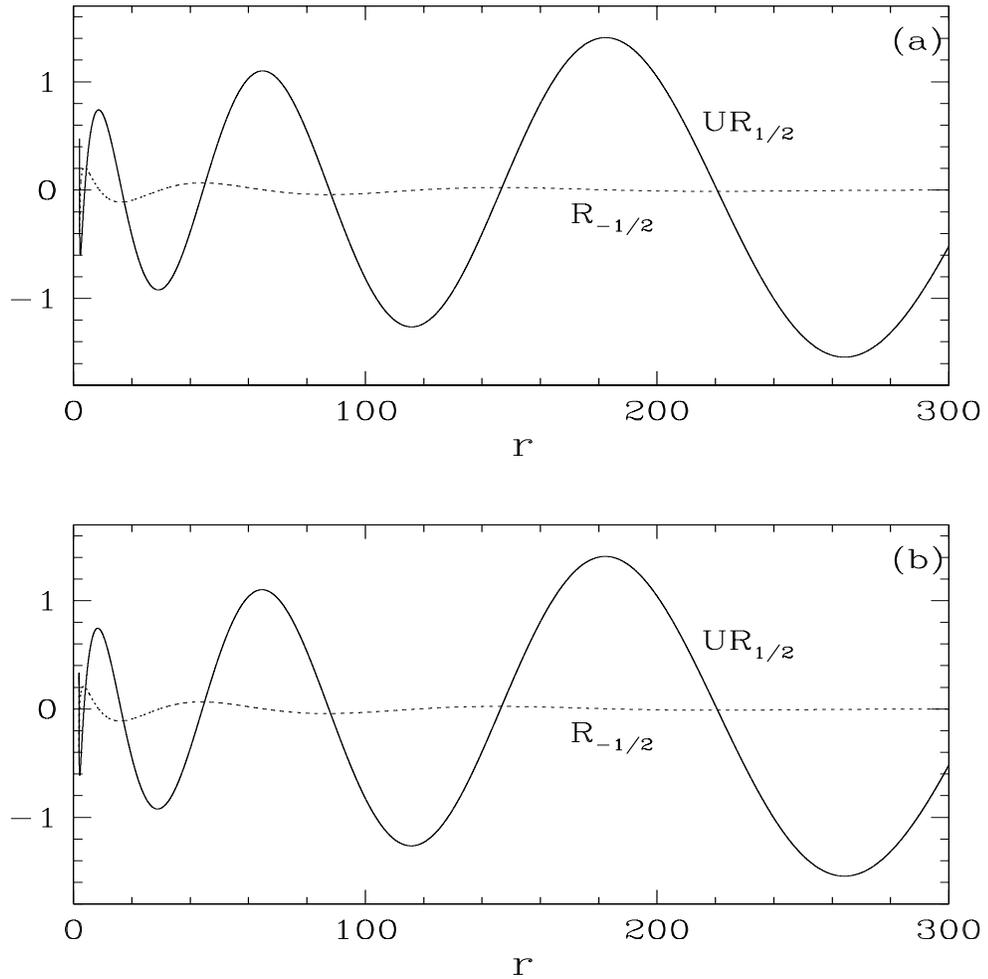,height=14truecm,width=14truecm,angle=0}}}
\vspace{-0.0cm}
\caption{ Radial spin-up and spin-down solutions with $\sigma=m_p=0.4$,
for (a) Kerr-NUT, $a=l=0.5$, (b) Kerr, $a=0.5$. 
Solid and dotted curves indicate the up and down spinors respectively as marked in the
figure too. Other parameters are $m=-1/2$, $Q_*=0$, $M=1$.
}
\label{fig14}
\end{figure}

Figures \ref{fig13} and \ref{fig14} depict the corresponding radial evolution of
spinors. The opposite sign of spin-spin coupling between the black hole and spinors of up and
down alignment reflects into the solutions which show the amplitude of the spin down wave is 
significantly
less compared to that of the spin up. The solutions in the dual Kerr (Fig. \ref{fig13}c) are 
shown in the semi-log scale
just to catch the abrupt variations of curves as $r\rightarrow 0$ in the same plot extended upto a
large $r$. At $r\rightarrow\infty$, $V_\pm\rightarrow m_p^2$, therefore the wave number 
of spinors becomes small at far away from the event horizon which produces a 
large wavelength compared to that at a small $r$.
For $\sigma=0.4$, cases with $a=0.5$ lie in the super-radiance regime for the dual Kerr
and then the corresponding radial potentials diverge.

\begin{figure}
\vbox{
\vskip -0.5cm
\hskip 0.0cm
\centerline{
\psfig{figure=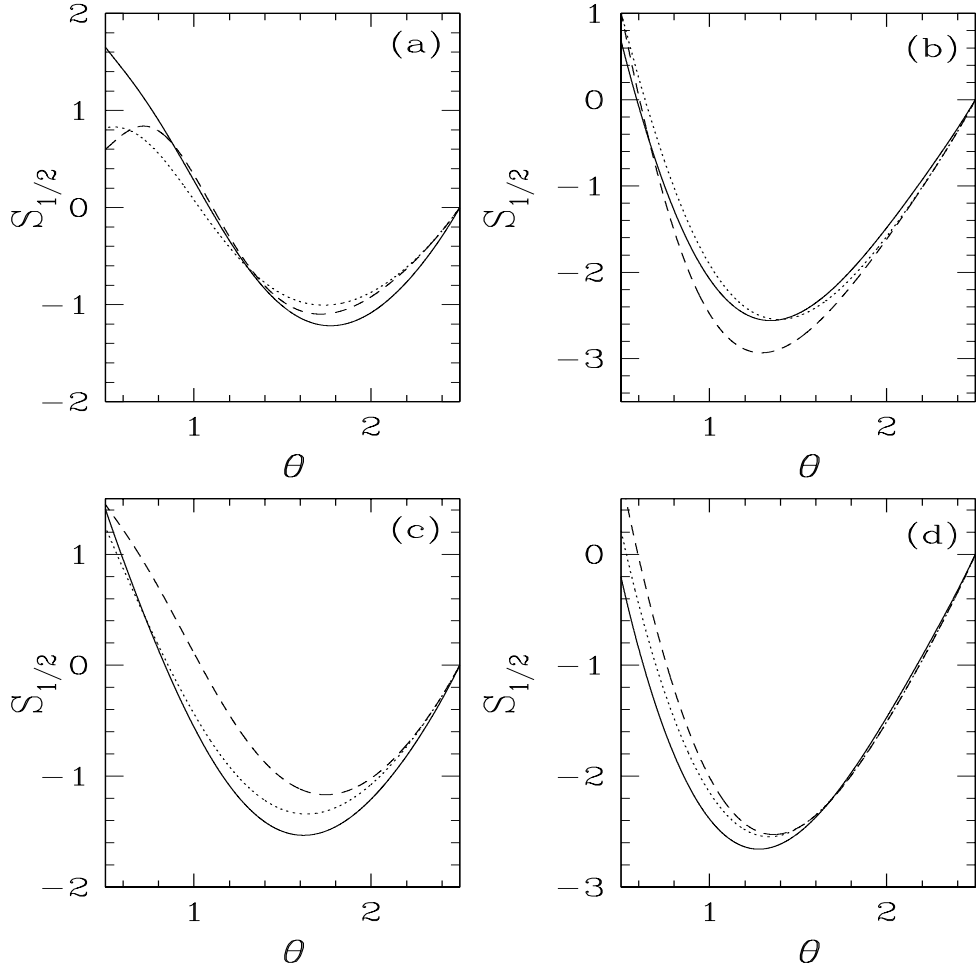,height=14truecm,width=14truecm,angle=0}}}
\vspace{-0.0cm}
\caption{ Angular spin-up solutions with $\sigma=0.4$, $m_p=0.1$ (solid curve); $\sigma=0.4$, $m_p=0.4$ 
(dotted curve); $\sigma=0.7$, $m_p=0.4$ (dashed curve);
for (a) Kerr-NUT, $a=0.998$, $l=0.99$, (b) Kerr, $a=0.998$, 
(c) Kerr-NUT, $a=l=0.5$, (d) Kerr, $a=0.5$.
Other parameters are $m=-1/2$, $Q_*=0$.
}
\label{fig15}
\end{figure}

\begin{figure}
\vbox{
\vskip -0.5cm
\hskip 0.0cm
\centerline{
\psfig{figure=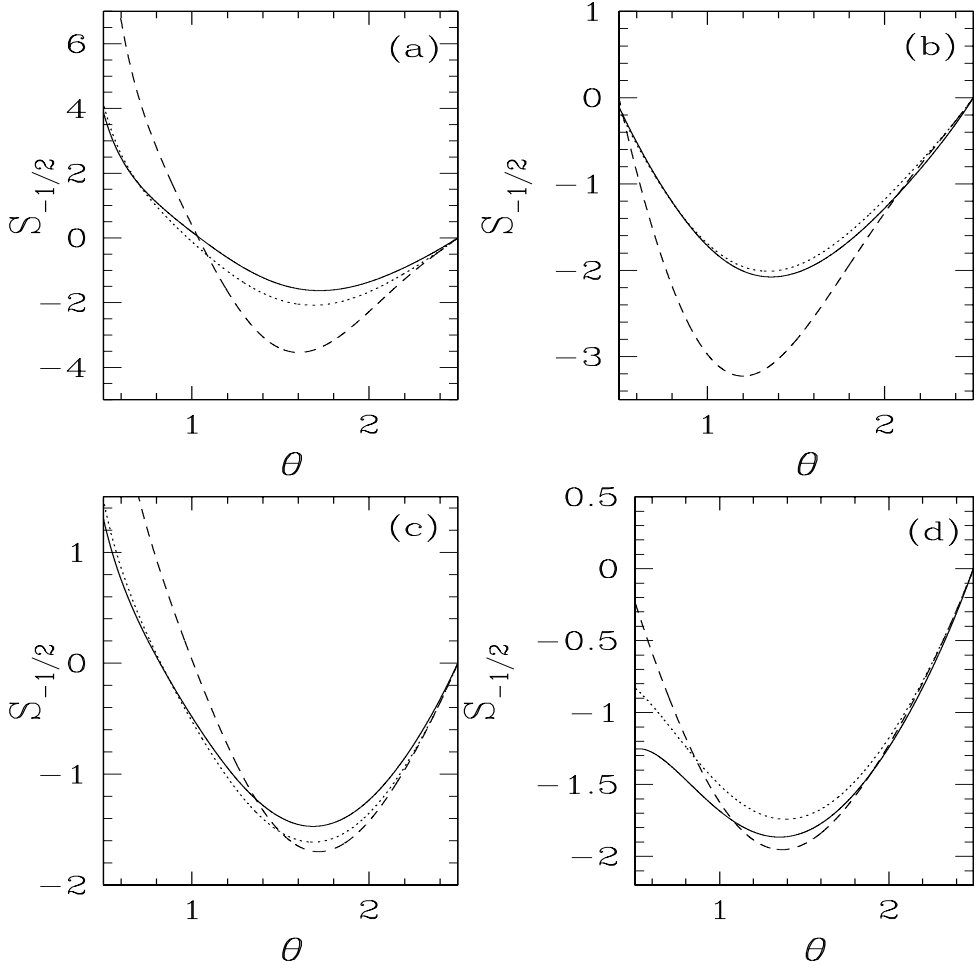,height=14truecm,width=14truecm,angle=0}}}
\vspace{-0.0cm}
\caption{ Angular spin-down solutions with $\sigma=0.4$, $m_p=0.1$ (solid curve); $\sigma=0.4$, $m_p=0.4$ 
(dotted curve); $\sigma=0.7$, $m_p=0.4$ (dashed curve);
for (a) Kerr-NUT, $a=0.998$, $l=0.99$, (b) Kerr, $a=0.998$, 
(c) Kerr-NUT, $a=l=0.5$, (d) Kerr, $a=0.5$.
Other parameters are $m=-1/2$, $Q_*=0$.
}
\label{fig16}
\end{figure}

Now we will compare the solutions of the spinor perturbation equations in the various regimes of 
incoming spinors so that the black hole will appear to act as a mass-spectrograph. 
Figures \ref{fig15} and \ref{fig16} bring this feature for the angular up and down spinors
respectively. As before only that range of $\theta$ is chosen when the solution is harmonic
and it however covers most of the range of $\theta$. The deviation of solutions for different 
sets of $\{\sigma,m_p\}$ is very clearly shown in the figures for the Kerr and Kerr-NUT space-times.

\begin{figure}
\vbox{
\vskip -0.5cm
\hskip 0.0cm
\centerline{
\psfig{figure=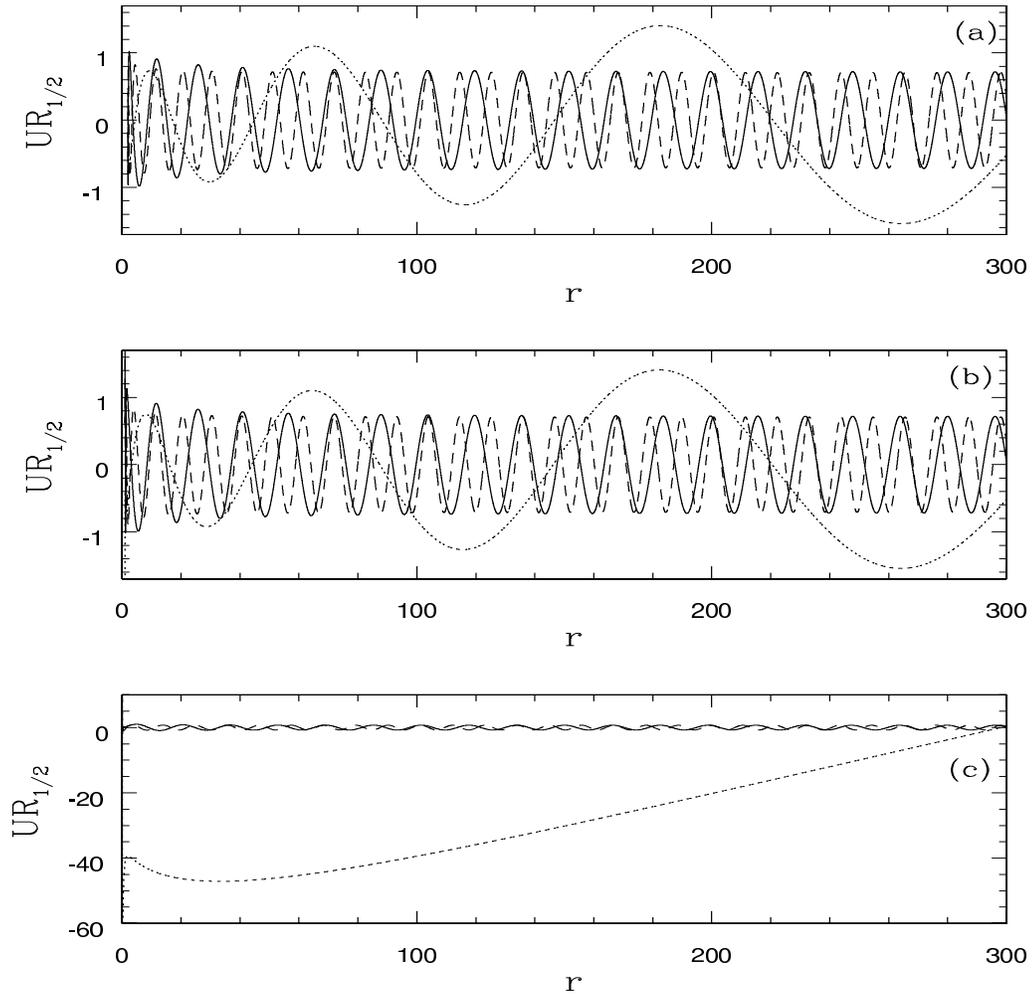,height=14truecm,width=14truecm,angle=0}}}
\vspace{-0.0cm}
\caption{ Radial spin-up solutions with $\sigma=0.4, m_p=0.1$ (solid curve); $\sigma=0.4, m_p=0.4$ (dotted curve);
$\sigma=0.7, m_p=0.4$ (dashed curve);
for (a) Kerr-NUT, $a=0.998$, $l=0.99$ (b) Kerr, $a=0.998$, (c) dual Kerr, $a=l=0.998$.
Other parameters are $m=-1/2$, $Q_*=0$; for (a) and (b) $M=1$.
}
\label{fig17}
\end{figure}

\begin{figure}
\vbox{
\vskip -0.5cm
\hskip 0.0cm
\centerline{
\psfig{figure=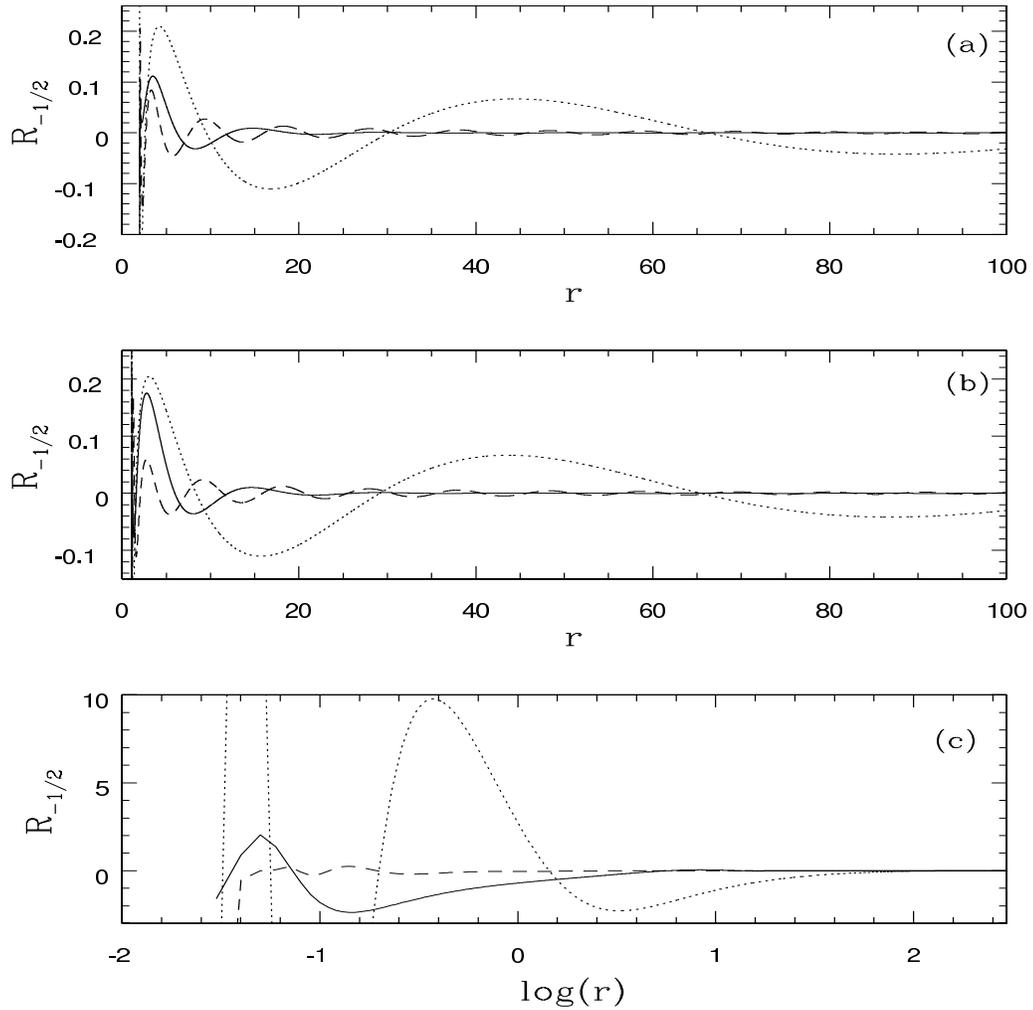,height=14truecm,width=14truecm,angle=0}}}
\vspace{-0.0cm}
\caption{ Radial spin-down solutions with $\sigma=0.4, m_p=0.1$ (solid curve); $\sigma=0.4, m_p=0.4$ (dotted curve);
$\sigma=0.7, m_p=0.4$ (dashed curve);
for (a) Kerr-NUT, $a=0.998$, $l=0.99$ (b) Kerr, $a=0.998$, (c) dual Kerr, $a=l=0.998$.
Other parameters are $m=-1/2$, $Q_*=0$; for (a) and (b) $M=1$.
}
\label{fig18}
\end{figure}

Figures \ref{fig17} and \ref{fig18} indicate the solutions for the up and down spinors respectively
in the radial direction
in case of the extreme Kerr-NUT, Kerr and dual Kerr space-times for different choices of $\{\sigma,m_p\}$.
The most interesting fact in all the figures is that when $\sigma\sim m_p$ (dotted curves)  
the wavelength as well as the 
amplitude of the spinors are high compared to two other cases. This is physically understood as $\sigma\sim m_p$ 
indicates the very small wave number at least at a large distance, that trend continues up to 
the inner radii, except
at very close to the horizon where the strong effect of a black hole suppresses any individual identity of
the matter. For $\sigma >m_p$, the perturbation of higher frequency gives rise to
the solution of
lower wavelength which is physically understood.
Again the solutions for the dual Kerr are distinctly different from others. As before the overall amplitudes
of up spinors which are oriented with the black hole's spin in a parallel sense, are high compared to that 
of down spinors oriented in an anti-parallel sense. 
 
\begin{figure}
\vbox{
\vskip -0.5cm
\hskip 0.0cm
\centerline{
\psfig{figure=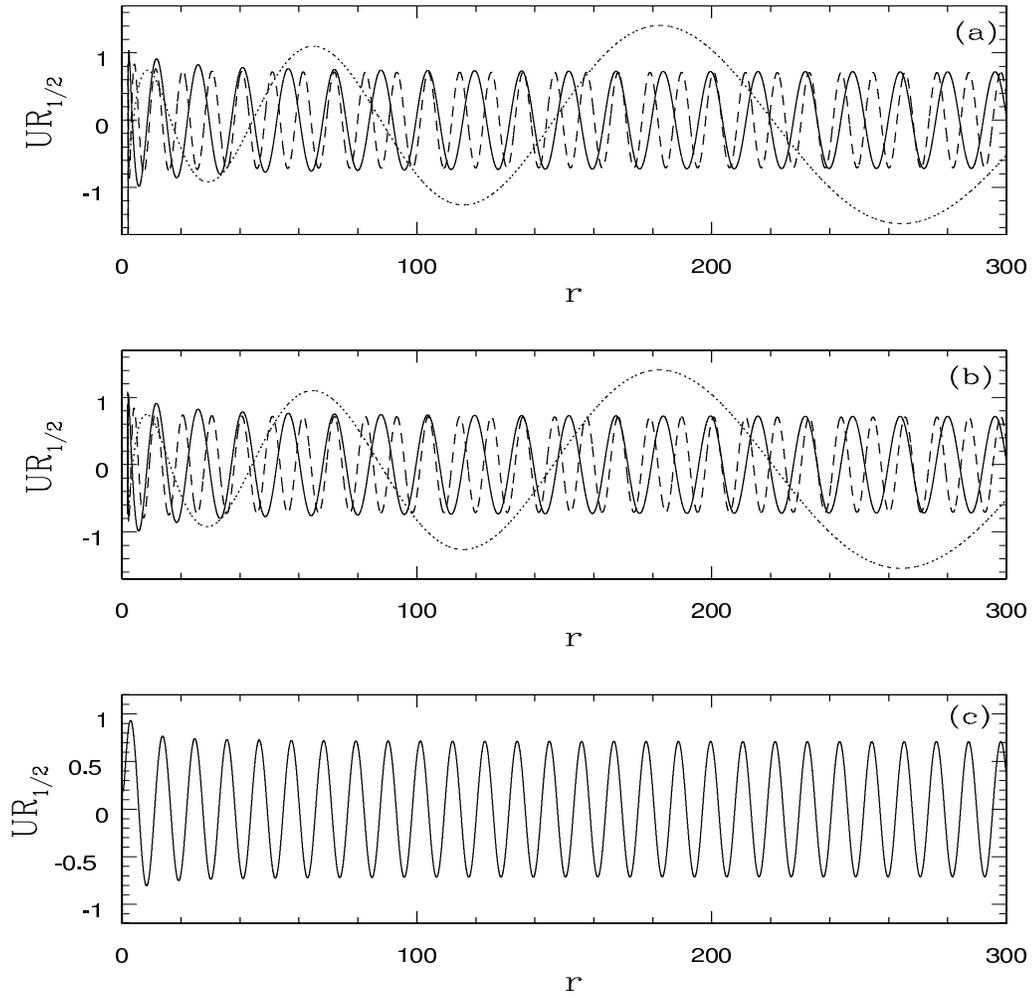,height=14truecm,width=14truecm,angle=0}}}
\vspace{-0.0cm}
\caption{ Radial spin-up solutions with $\sigma=0.4, m_p=0.1$ (solid curve); $\sigma=0.4, m_p=0.4$ (dotted curve);
$\sigma=0.7, m_p=0.4$ (dashed curve);
for (a) Kerr-NUT, $a=l=0.5$ (b) Kerr, $a=0.5$, (c) dual Kerr, $a=l=0.5$.
Other parameters are $m=-1/2$, $Q_*=0$; for (a) and (b) $M=1$.
}
\label{fig19}
\end{figure}

\begin{figure}
\vbox{
\vskip -0.5cm
\hskip 0.0cm
\centerline{
\psfig{figure=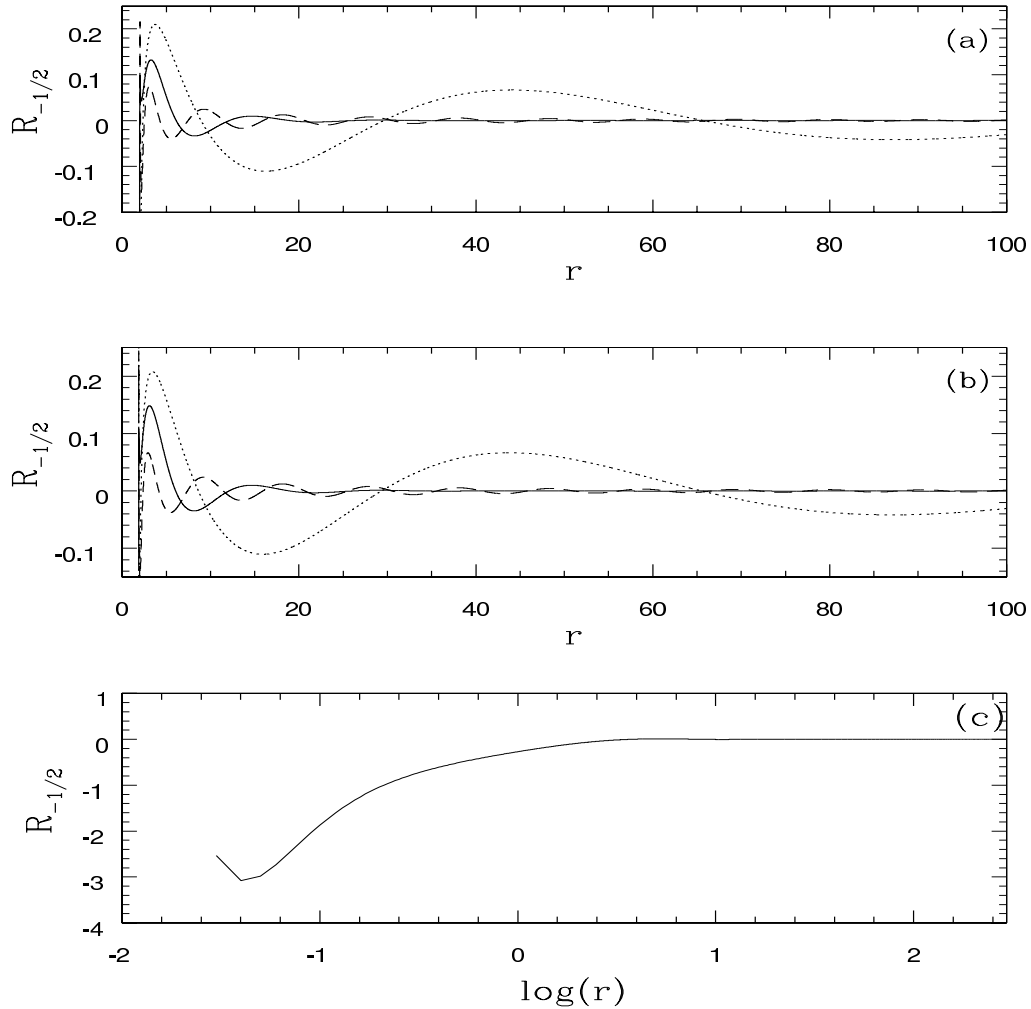,height=14truecm,width=14truecm,angle=0}}}
\vspace{-0.0cm}
\caption{ Radial spin-down solutions with $\sigma=0.4, m_p=0.1$ (solid curve); $\sigma=0.4, m_p=0.4$ (dotted curve);
$\sigma=0.7, m_p=0.4$ (dashed curve);
for (a) Kerr-NUT, $a=l=0.5$ (b) Kerr, $a=0.5$, (c) dual Kerr, $a=l=0.5$.
Other parameters are $m=-1/2$, $Q_*=0$; for (a) and (b) $M=1$.
}
\label{fig20}
\end{figure}

The similar sets of solution are compared in Figs. \ref{fig19} and \ref{fig20} but for an intermediate
rotating black hole. All the basic features are similar to that of Figs. \ref{fig17} and \ref{fig18} 
which are
not repeated again. For the black hole angular momentum (Kerr parameter) $a=0.5$ the corresponding
$\sigma_s=0.5$ in case of the dual Kerr with $l=0.5$. Therefore only the $\sigma=0.7$ case is out of 
the super-radiance regime and depicted in Figs. \ref{fig19}c and \ref{fig20}c.

\section*{VI. Horizon and singularity}

The distinguishing feature of the Kerr-NUT geometry is the presence of the magnetic charge 
which is called the NUT
parameter, $l$, that causes the space-time to be asymptotically non-flat. The horizons are
defined by $U=0$ and hence they will occur at $r_\pm=M\pm\sqrt{M^2+l^2-a^2-Q_*^2}$. We shall now
switch off the electric charge, i.e. $Q_*=0$, as it is not very pertinent to carry it through for the purpose
in question. 
The singularity would occur where curvatures diverge and that would
happen when $r^2 + (l + acos\theta)^2 = 0$. That means singularity would be
located at $r = 0, cos\theta = -l/a$. Clearly there occurs no singularity
when $l > a$. It is the ring like ($r=0, \theta =\pi/2$) for the Kerr case
when $l = 0$ and string like when $a > l > 0$ (Kerr-NUT case). The string like
character of the singularity is the characteristic of the presence of the NUT
parameter \cite{si}.
                                                                                                 
On the other hand, horizon would always occur for $M^2 + l^2 \geq a^2$.
For $l > a$, there occurs an interesting situation of having a horizon
without singularity. The space-time is then non singular and regular everywhere.
This is very strange because horizon should generally cover singularity.
It has happened due to dominance of the NUT parameter over the Kerr
parameter. It is however well known that the NUT parameter also gives rise
to the closed time-like geodesics. This suggests that the NUT parameter
should never dominate over the Kerr parameter to ensure physical
reasonableness.

If however, we wish to have the canonical picture of a black hole space-time having the horizon covering a singularity,
then $l^2<a^2$, which would imply $a\neq0$ always. Recall that this was also required for the duality transformation between
the electric (mass) and magnetic (NUT) charge. In case of the dual Kerr solution $M=0$ and the horizon would be given by
$r=\pm\sqrt{l^2-a^2}$ which would require $l^2\geq a^2$, while existence of singularity would require $l^2\leq a^2$.
Requiring both the horizon and singularity to coexist, we are unambiguously led to $l^2=a^2$. The horizon occurs at
$r=0$ and it becomes singular when $\theta=0,\pi$. The dual Kerr solution may not be very realistic but it provides an
interesting vacuum space-time which is free of the usual electric gravitational charge (mass). The most remarkable feature
is that general relativity admits a truly gravitational dyon solution which has the magnetic limit as the dual Kerr
solution.

\section*{VII. Discussion}

We have studied the scalar and spinor perturbation in the Kerr-NUT 
(which includes the Kerr and dual Kerr) space-time which describe gravitational
field of a rotating gravitational dyon with electric and magnetic charge. It is known that the Kerr-NUT
space-time is invariant under the duality transformation which exchanges mass and NUT parameters
as well as radial and angle coordinates \cite{dt1,dt2}. By the duality transformation one can go
from the Kerr to the dual Kerr solution and the vice versa. Interestingly, this duality transformation requires the
rotation parameter to be non zero. Here we have shown that the same is true for the Klein-Gordon and Dirac
equations in the Kerr-NUT space-time. That is, they are invariant under the duality, and they transform
from the Kerr background to the dual Kerr, and the vice versa under the duality transformation with some appropriate rescaling of parameters.

In the Kerr geometry, angular part of the equation is free from mass and simply involves the kinematic
aspects arising from rotation. With the NUT parameter, angular part attains active dynamical meaning by
its presence. The radial part would however involve both the mass and NUT parameters. The behaviour of
potentials and their solutions in different cases are shown in various figures. Potential
barriers are higher for the dual Kerr space-time. This is because of the absence of mass, which produces
the usual $1/r$ attractive potential while the NUT parameter, $l$, contributes $1/r^2$ asymptotically. Thus gravity is as expected stronger for the Kerr than 
the dual Kerr.

Though the Klein-Gordon and Dirac equations could be transformed under the duality transformation from
the Kerr to the dual Kerr case, their solutions could not be transformed into each other by any
simple duality transformation. Had the duality worked for the solutions of the equations, 
it would have indicated something profound. The 
problem is that in the Kerr background $l=0$ while for the dual Kerr $M=0$, 
the character of the equation changes drastically in the two cases and hence their solutions can 
not be related by the duality transformation. This however does not rule out the possibility that 
there may occur some other transformation 
which may work with the solutions. Note that in obtaining the solutions in a known form, we have defined the new independent variables 
in a complicated form which does not let duality transformation work at the solution level. With a proper and suitable new definition, it may be possible to 
relate the solutions by the duality transformation. This as well as finding the complete solution 
with $\lambda_1, \lambda_2$ properly determined (rather than the qualitative solutions as considered 
here) of the perturbation equation we leave for future study. 
But we do however believe that this would not significantly alter the qualitative character of the solutions.

%\noindent{\large\bf Acknowledgment}\\

\begin{acknowledgements}

We thank the referees for the useful suggestions that have improved
the presentation of this paper. One of the author (B.M.) acknowledges
the partial support by NSF grant AST 0307433 and NASA grant NAG5-10780
to this work.
                                                                                
\end{acknowledgements}


\begin{references}



\bibitem {chandra} S. Chandrasekhar, in {\it The Mathematical Theory Of Black Holes}
(London: Clarendon Press, 1983).

 \bibitem {nut} E. T. Newman, L. Tamburino \& T. Unti, J. Math. Phys. {\bf 4}, 915 (1963).
 \bibitem {carter} B. Carter, Comm. Math. Phys. {\bf 10}, 280 (1968).
 \bibitem {new-dema} M. Demianski \& E. T. Newman, Bull. Acad. Polon. Sci. {\bf 14}, 653 (1966).

  \bibitem {dt1} N. Dadhich \& Z. Ya. Turakulov, Class. Quantum Grav. {\bf 19}, 2765 (2002).

  \bibitem {dt2} N. Dadhich \& Z. Ya. Turakulov, Mod. Phys. Lett. {\bf A 17}, 1091 (2002).
  \bibitem {ly-no} D. Lynden-Bell \& M. Nouri-Zonoz, Rev. Mod. Phys. {\bf 70}, 427 (1998).
  \bibitem {dp} N. Dadhich \& L. K. Patel, J. Math. Phys. {\bf 41}, 882 (2000). 
  \bibitem {td} Z. Ya. Turakulov \& N. Dadhich, Mod. Phys. Lett. {\bf A 16}, 1959 (2002).

   \bibitem {teu1} S. Teukolsky, Phys. Rev. Lett. {\bf 29}, 1114  (1972).

   \bibitem {chandra1} S. Chandrasekhar, Proc. Roy. Soc. Lond. {\bf A 349}, 571 (1976).

    \bibitem {mc99} B. Mukhopadhyay \& S. K. Chakrabarti, Class. Quantum Grav. {\bf 16}, 3165 (1999).

     \bibitem {mc00} B. Mukhopadhyay \& S. K. Chakrabarti, Nucl. Phys. {\bf B 582}, 627 (2000).


      \bibitem {m00} B. Mukhopadhyay, Class. Quantum Grav. {\bf 17}, 2017 (2000).

       %\bibitem {} S. Teukolsky, Astrophys. J. {\bf 185}, 635 (1973).

        \bibitem {pt} W. H. Press \& S. Teukolsky, Astrophys. J. {\bf 185}, 649 (1973).

         \bibitem {tp} S. Teukolsky \& W. H. Press, Astrophys. J. {\bf 193}, 443 (1974).

%         \bibitem {bl}



                  \bibitem {chak} S. K. Chakrabarti, Proc. Roy. Soc. Lond. {\bf A 391}, 27 (1984).


                  \bibitem {wd} S. M. Wagh \& N. Dadhich, Phys. Rev. {\bf D 32}, 1863 (1985).

                  \bibitem {si} J. Samuel \& B. R. Iyer, Current Science {\bf 55}, 818 (1986).




\end{references}
\end{document}